\documentclass[12pt,preprint]{aastex}
\usepackage{times,pslatex}
\usepackage{amsmath,amsfonts,graphicx}

\newcommand{\vRstarA}{1.7}
\newcommand{\eRstarA}{0.06}
\newcommand{\vRstarB}{0.34}
\newcommand{\eRstarB}{0.015}
\newcommand{\vRplanet}{0.52}
\newcommand{\eRplanet}{0.018}

\newcommand{\vMstarA}{1.47}
\newcommand{\eMstarA}{0.08}
\newcommand{\vMstarB}{0.37}
\newcommand{\eMstarB}{0.035}

\newcommand{\kps}{km~s$^{-1}$}
\newcommand{\Mjup}{\mbox{$M_{\rm J}$}}

\newcommand{\Msun}{\mbox{$M_\odot$}}
\newcommand{\Rsun}{\mbox{$R_\odot$}}
\newcommand{\Mpri}{\mbox{$M_{\rm pri}$}}
\newcommand{\Msec}{\mbox{$M_{\rm sec}$}}
\newcommand{\Rpri}{\mbox{$R_{\rm pri}$}}
\newcommand{\Rsec}{\mbox{$R_{\rm sec}$}}

\newcommand{\Vsini}{\mbox{$v\sin i_*$}}

\slugcomment{Accepted for publication in the Astrophysical Journal}

\received{2012 Oct 14}
\begin{document}

\title{A Gas Giant Circumbinary Planet Transiting the F Star Primary of the Eclipsing Binary Star KIC 4862625 and the Independent Discovery and Characterization of the two transiting planets in the Kepler-47 System.}
\shorttitle{Circumbinary Planets KIC 4862625b and Kepler-47bc}

\author{
V.~B.~Kostov\altaffilmark{1,2},
P.~R.~McCullough\altaffilmark{2},
T.~C. Hinse\altaffilmark{3,4},
Z.~I.~Tsvetanov\altaffilmark{1},
G.~H\'{e}brard\altaffilmark{5,6},
R.~F.~D\'{\i}az\altaffilmark{7},
M.~Deleuil\altaffilmark{7},
J.~A.~Valenti\altaffilmark{2}
}

\email{vkostov@pha.jhu.edu}

\altaffiltext{1}{Department of Physics and Astronomy, Johns Hopkins University, 3400 North Charles Street, Baltimore, MD 21218}
\altaffiltext{2}{Space Telescope Science Institute, 3700 San Martin Dr., Baltimore MD 21218}
\altaffiltext{3}{Korea Astronomy
and Space Science Institute (KASI), Advanced Astronomy and Space Science Division, Daejeon 305-348, Republic of Korea}
\altaffiltext{4}{Armagh Observatory, College Hill, BT61 9DG Armagh, NI, UK}
\altaffiltext{5}{Institut dÕAstrophysique de Paris, UMR 7095 CNRS, Universit\'{e} Pierre \& Marie Curie, 98bis boulevard Arago, 75014 Paris, France}
\altaffiltext{6}{Observatoire de Haute-Provence, Universit\'{e} dÕAix-Marseille \& CNRS, 04870 Saint Michel lÕObservatoire, France}
\altaffiltext{7}{Laboratoire d'Astrophysique de Marseille, 38 rue Fr\'ed\'eric Joliot-Curie, 13388 Marseille cedex 13, France}

\begin{abstract}
We report the discovery of a transiting, gas giant circumbinary
planet orbiting the eclipsing binary \object{KIC 4862625} and describe our
independent discovery of the two transiting planets orbiting Kepler-47
(Orosz et al. 2012). We describe a simple and semi-automated procedure
for identifying individual transits in light curves and present our
follow-up measurements of the two circumbinary systems.  
For the \object{KIC 4862625} system, the $\vRplanet
\pm \eRplanet\ R_{Jupiter}$ radius planet revolves every $\sim
138$ days and occults the $\vMstarA \pm \eMstarA\ \Msun$, $\vRstarA
\pm \eRstarA\ \Rsun$ F8 IV primary star producing
aperiodic transits of variable durations commensurate with the
configuration of the eclipsing binary star. Our best-fit model indicates the orbit has
a semi-major axis of $0.64$ AU and is slightly eccentric, $e=0.1$.
For the Kepler-47 system, we confirm the results of Orosz et al. (2012).
Modulations in the radial velocity of \object{KIC 4862625A}
are measured both spectroscopically and photometrically, i.e. via Doppler
boosting, and produce similar results.
\end{abstract}


\include{variables}
\section{Introduction}
\label{sec:intro}

For decades the science fiction community has imagined that planets can orbit binary stars, yet only recently have such systems actually been detected. Timing variations either in the rotation period of a neutron star member of a binary system (Sigurdsson et al., 2003) or in the stellar occultations (when the two stars eclipse each other) of eight eclipsing binary (EB) systems (Deeg et al. 2008; Lee et al. 2009; Beuermann et al. 2010, 2011; Potter et al. 2011; Qian et al. 2011; Qian et al. 2012) have been interpreted as the gravitational perturbation of additional bodies on the binary stellar system, suggesting the presence of a total of 12 circumbinary (CB) planets on wide orbits with periods of tens of years. 

The lower limits on the masses of all twelve objects, however, fall in the super-Jupiter regime, making their planetary nature uncertain. Furthermore, the orbital stability of some of the multi-planet circumbinary systems (HW Vir, HU Aqr and NN Ser) have been studied recently, showing that some of them are on highly unstable orbits \citep{Horner2011, Hinse2012, Horner2012a, Horner2012b, Gozdziewski2012, Beuermann2012}.

Doyle et al. (2011) announced the first direct evidence of a Saturn-sized planet that transits both members of an EB eclipsing binary, specifically Kepler-16.
Since then, five more CB planets have been announced: Saturn-sized planets that transit Kepler-34b and Kepler-35b (Welsh et al. 2012), the first Neptune-sized planet that transits Kepler-38 (Orosz et al. 2012b), and two Neptune-sized planets that transit Kepler-47 (Orosz et al. 2012). The latter system is the first binary discovered to have two planets, one Neptune-sized on a 300 day orbit and the other Earth-sized on a ~49.5 day orbit. 

Substantial efforts in theoretical modeling indicate that planets such as these should not be uncommon. Simulations of dynamical stability show that beyond a critical distance, CB planets can have stable orbits in practically all binary configurations. The critical distance is on the order of a few binary separations (Dvorak 1986; Holman \& Weigert 1999; Scholl et al. 2007; Haghighipour et al. 2010; Schwarz et al. 2011; Doolin \& Blundell 2011). The orbits of Kepler-16b, 34b, 35b and 38b are indeed outside the critical orbital semi-major axis, but only by 21\%, 24\%, 14\% and 26\% respectively (Welsh et al. 2012; Orosz et al. 2012b). Kepler-47b, while notably farther from the instability region (Orosz et al. 2012), is still not too far out, suggesting such ``reaching for the limit'' behavior to be typical of CB planets. 

The fact that these four planets are so close to the theoretical limit for stability may suggest that their host systems had an interesting dynamical history where migration and/or planet-planet scattering may have played a significant role in sculpting their present architecture. Formation and evolution theory of giant planets around binary stars has been studied extensively (Pierens and Nelson 2007, 2008a, 2008b), providing a number of outcomes that depend on initial conditions. Simulations of gas giants have shown that Saturn-size planets (like Kepler-16b, 34b and 35b) stabilize at a 5:1 orbital resonance and may be very common, compared to Jupiter-size planets that are either scattered out of the system or gradually drift outward into the disk. Single, Neptune-size planets (such as Kepler-38b) migrate and stop at a distance of about three times the binary stars separation, leading the authors to suggest that ``the cavity edge of the precursor CB disk appears to be an excellent place to look for low mass planets in close binary systems.'' If there are two Neptune-size planets in the system, they become locked in a mean motion resonance, while a five-planet system is either disrupted or, in one simulation, also ends up in a resonance, implying that such multiple Neptune-size planets in resonant orbits may be indeed common. Models for the formation and evolution of terrestrial planets around binary stars (Quintana \& Lissauer, 2006) have shown that, in the presence of Jupiter at 5 AU, CB terrestrial planets can readily form around a wide variety of binary systems. At least one terrestrial planet forms in all simulations presented by the authors. While the final masses of all simulated terrestrial planets vary little, the outcome for the architecture of the planetary system is very dependent on the parameters of the stellar binary. Highly eccentric binaries tend to harbor fewer, more diverse suite of planets compared to binaries with very low eccentricity, a prediction that can be addressed by the addition of more pictures to the family portrait of the five Kepler planets. 

More than 20 years ago Borucki and Summers (1984) proposed monitoring EB systems to search for planets because nearly edge-on inclination significantly increases the probability of transits. At the time it was not practical to monitor targets continuously over many days (Kepler-16b, for example, has an orbital period of 230 days). On March 6 of 2009, 380 years after Johannes Kepler predicted the transit of Venus across the disk of the Sun, NASA launched the appropriately named Kepler Mission to search for Earth-like planets in the habitable zone of Solar-type stars and to determine their occurrence frequency (Borucki et al. 2010). To achieve this, a 0.95 Schmidt telescope on a Heliocentric Earth trailing orbit continuously and simultaneously monitors about 150,000 stars in the visible range from 423~nm to 897~nm over a 100 square degrees patch of the sky in the Cygnus region where $\sim60\%$ of the stars are G-type stars on or near the Main Sequence. Utilizing the transit method, the instrument searches for periodic dips in the brightness of a star caused by a planet transiting across its disk. The Kepler mission has been remarkably successful in finding transiting planets, discovering more than 2300 planet candidates (Borucki et al. 2011; Batalha et al. 2012), 77 of which have been confirmed by the time of writing. Amongst this treasure throve of data are also a set of 2165 EB systems (Slawson et al. 2011), the main focus of our work. 

Searching light curves of EB stars for transits of a third body is non-trivial. In addition to the significant limitations associated with intrinsic stellar variability and instrumental artifacts, CB planets have {\it inconstant} transit times, durations, and depths, all of which that depend on the phase of the binary system (Schneider \& Chevreton 1990).  To transit one of the stars, the planet must ``hit a moving target'' (Orosz et al. 2012).
A benefit is that these transit signatures cannot be attributed to the stars themselves, to a background EB, or to other unrelated astrophysical or instrumental events, strongly supporting the CB-planet hypothesis. One challenge is that traditional transit searching algorithms, e.g. Box-fitting Least Squares (BLS) (Kovacs et al. 2002), used to detect periodic, box-like signals in the light curve of a single star are not optimized for finding transiting CB planets due to the unique nature of their signal. Several methods for the detection of transiting CB planets have been proposed. One approach is based on simulating light curves produced by an exhaustive search of possible orbits of CB-planets and fitting them to the data (Doyle et al. 2000; Ofir 2008). Carter \& Agol (2013) have developed the Quasi-periodic Automated Transit Search QATS algorithm, which is similar to BLS but optimized for aperiodic pulses. QATS has been successfully applied to CB-planets. Orosz et al. (2012) report that QATS failed to detect the outer planet Kepler-47c due to decreasing sensitivity for longer periods. While transiting gas giants cause a dimming of their host star large enough to be seen by eye in light curves, the transits of smaller planets can be easily missed by visual inspection. 

The initial discovery of CB transits together with the availability of exquisite Kepler data inspired us to develop a semi-automated procedure to identify aperiodic transits. 
We describe the procedure, which is based on the established BLS algorithm but modified and applied in a novel way. 
We applied it to finding transiting planets around EB stars listed in the Kepler catalog of Slawson et al. (2011). We examined the detached EB systems and identified several candidates that exhibited additional transit-like features in their light curves. 
Here we present the independent discoveries of two CB planets \object{Kepler-47bc} and \object{KIC 4862625b}.

In this paper, we will describe the analysis as a linear, deductive process, although it is inherently iterative, with one aspect feeding back into an earlier part. For brevity and clarity, we do not emphasize the iterations. This paper is organized as follows. In Section \ref{sec:photometry} we describe the procedure used to discover the two CB systems, followed by radial velocity measurements and spectra of the host stars in Section \ref{sec:spectra}. Our data analysis and initial diagnosis of the \object{KIC 4862625} system are outlined in Sections \ref{sec:data} and \ref{sec:puzzle} respectively. We present our results in Section \ref{sec:results} and describe the dynamical stability of the \object{KIC 4862625} system in Section \ref{sec:dynamics}. In Section \ref{sec:discussion} we present our discussion and draw conclusions in Section \ref{sec:conclusions}. 
\section{Kepler Photometry}
\label{sec:photometry}

\subsection{Detrending}
\label{sec:detrending}

We began with the long-cadence ($\sim$30 min) PDCSAP flux of \object{KIC 4862625}
generated by the Kepler mission for the publicly available\footnote{http://archive.stsci.edu/kepler/ or http://exoplanetarchive.ipac.caltech.edu .} quarters 1-14.
Using the ephemeris of Pr{\v s}a et al. (2010) and examining the phase-folded light curve, we flagged
data points within 0.12 days of the centers of the primary or secondary eclipses,
or within 0.5 days of the planetary transit events.\footnote{This is one example of the iterative analysis: first we detrended the light curve, then we identified the planetary transits, then we detrended the light curve again with the transit points flagged.} To remove the instrumental discontinuities in flux created by the quarterly rotation of the Kepler focal plane, we divide each quarter's data by its median.
We flagged points that differed from the rest by more than 0.2\% in normalized flux.

\begin{figure}
\centering
\plotone{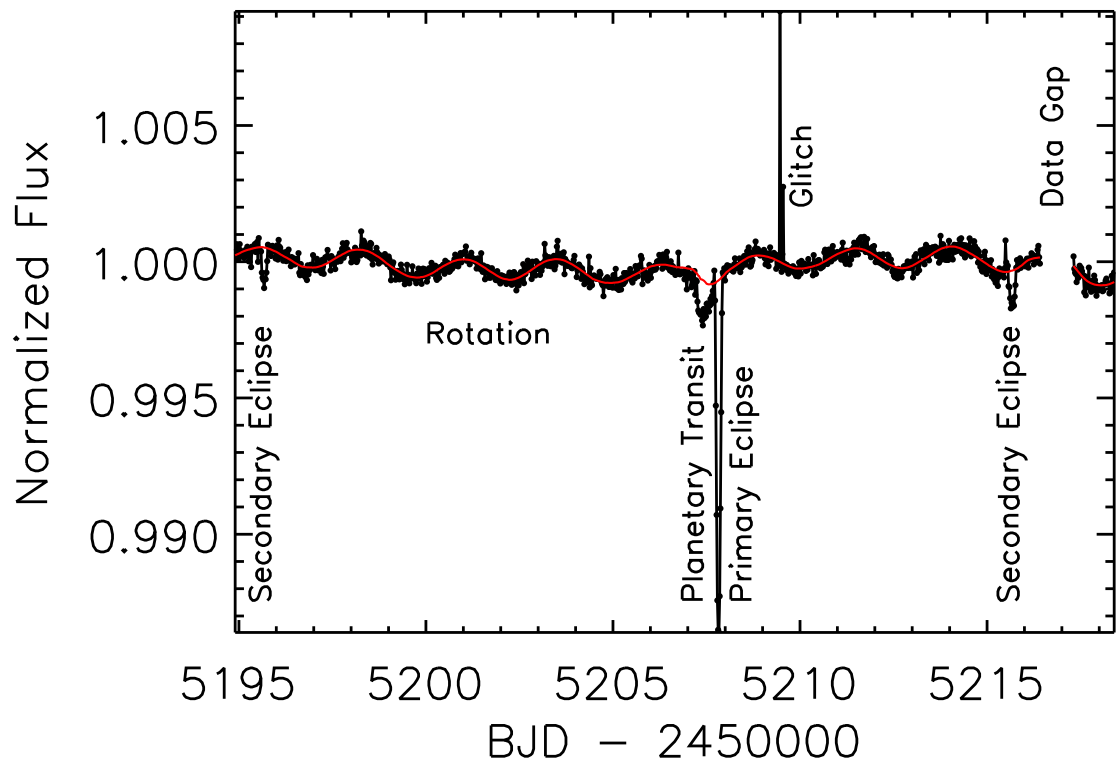}
\caption{
Characteristics of the light curve of {\protect \object{KIC 4862625}}. A 23-day portion of the Kepler light curve illustrates 
various phenomena. Instrumental effects are labeled above the data; astrophysical effects are
labeled below the data. We flagged two data points as a ``glitch,'' and a gap
in the data is visible at BJD = 2455217. 
The eccentric orbit of the EB is apparent
from the fact that primary eclipse is not centered in time between the two secondary eclipses.
The primary eclipse is $\sim 1.4$\% deep. 
The red line illustrates the $\sin$-wave fit to the rotational modulation of the light curve.
The secondary eclipses, the planetary transit, and the
2.6-day rotational modulation all have similar amplitudes, $\sim 0.1$\%. 
\label{fig:typicallc}}
\end{figure}

The unflagged flux is sinusoidal with a period of ${\rm p_{rot} \approx 2.63}$ d, which
we attribute to modulation caused by the rotation of star A.
To detrend the flux variations attributed to rotation, for the purposes of providing 
eclipse and planetary transit light curves normalized to unity,
we fit a unique $\sin$ wave in the local vicinity of each data point (58763 in total), in each case using 
unflagged data within ${\rm p_{rot}/2}$ of each data point. Each of the 58763 $\sin$ waves had a fixed period, 
${\rm p_{rot}}$, and we fit for three parameters: the mean, the amplitude, and the phase.
We iterated the procedure, adjusting the value of ${\rm p_{rot}}$, until the gradient in the phase shifts of the
fitted $\sin$ waves over the entire data span was zero, indicating a best-fitted average rotational period,
${\rm p_{rot} = 2.6382\pm 0.0037}$ d, where the quoted ``uncertainty'' is the standard
deviation of ${\rm p_{rot}}$ fit piece-wise over quarterly time spans.
Finally, we divided each data point by its best-fitting sine wave evaluated at the time of each particular
data point. A small section of the light curve illustrating various features (stellar eclipses and planetary transit, rotational modulation, data gaps and glitches) is shown on Fig. \ref{fig:typicallc}. The part near the planetary transit where the best-fit $\sin$ wave deviates from the rest is caused by an instrumental effect, present in many other light curves. It does not affect the determination of the primary eclipse depth.
We use the resulting detrended, normalized light curve for subsequent analysis, and we
use the mean values of the $\sin$ waves for the analysis of the Doppler boosting (Section \ref{sec:boosting}).
Exclusive of flagged data, the RMS of the residuals of the detrended, normalized light curve about unity
is 222 ppm.

\subsection{Box-fitting Least Squares, BLS, for Single Transits}
\label{sec:bls}

We invented a semi-automated procedure to identify individual,
aperiodic transit-like features in a light curve.  The procedure
automatically finds square-wave or ``box-shaped'' features within
a light curve. 

Using the raw EB light curves SAPFLUX from the Kepler database we first normalize them and remove the EB's eclipses using the Box-fitting Least Squares method (BLS, Kov{\'
a}cs et al. 2002). Next we detrend the data, using only those points with a SAPQUALITY flag of 0. The detrending is non-trivial and has to be done on a target-by-target basis, as each binary star has complex baseline variability that spans timescales from hours to days. We use an iterative fit with a high-order Legendre polynomial on each quarter, further broken down into smaller segments according to the data flags.

Within each segment of the light curve, the procedure automatically identifies the center,
width, and depth of the two most-significant box-shaped features,
one positive and one negative. The latter is a ``transit''
candidate and the former is an ``anti-transit.''
Because transits are negative features in the residuals
of a detrended light curve, we can validate empirically the statistical
significance of the transit candidates by comparison with the 
anti-transits extracted from the same data. Doing so helps us to not be
overwhelmed with false-positives.

The number M of light-curve segments is not particularly critical. The segment length
should be longer than any actual transit.  For CB planets, in
principle the transits can be as long as half an orbit of the EB,
in the case where the planet and the star are traveling in parallel at nearly
the same projected rate (Schneider \& Chevreton 1990). Such transits will be rare
and generally accompanied by shorter transits, created when the planet and star
are moving in opposite directions. The segment length should be shorter than the
orbital period of the planet, lest only one of two genuine transits be identified
in a segment. For CB planets of P-type (Dvorak 1982), the planet's orbital period must be
longer than the period of the EB, and for orbital stability the
CB planet's period must be at least a few times longer than the EB's period (Section \ref{sec:dynamics}).
From these two limits, the segment length should be between one half and a few times 
the orbital period of the EB. However, in practice, the intrinsic variability of
the stars in the system prescribe the length of the segment.
In any case, to prevent a transit being split by the segment
boundaries, we analyze each light curve at least twice with the boundaries of
the segments shifted. 

In all cases, detrending of the light curve is crucial to our method. Very short-period EB stars, contact, and semi-detached EBs are difficult to study with our method. They are highly variable and there are few
measurements in between the stellar eclipses, forcing us to use segments much larger than the binary period. 

To avoid systematics effects that might mimic a transit, a merit criterion is necessary.
As a convenience, one can use the BLS algorithm to find the individual transit- and
anti-transit-features in segments of the detrended light curve's residuals.  Because BLS
is designed to find {\it periodic} box-like features, one option is to replicate N
times the segment under study to form a periodic light curve.  One
can then use BLS to search for the most significant transit- and anti-transit
features with that single period defined by the replication.  The number N
of replications is not particularly important.  To find the anti-transits,
we simply invert the sign of the residuals of the light curve and
search a second time. Although it would be more efficient to
extract the most-significant positive and negative features in one pass, the execution
time is trivial and enormously smaller than in the traditional BLS
which must loop over thousands of trial periods, and smaller than the detrending also.
Alternatively one can modify the original BLS computer code to analyze a single segment
of unreplicated data; we have implemented each variant. The only
potential difference is a tuning parameter within BLS, namely the
minimum number of data points within an acceptable box-like feature,
which must be adjusted commensurate with the number of replications N.

After the automated procedure identifies the M most-significant
pairs of transit- and anti-transit-candidates from the M segments, we examine the ensemble
for outlier transit-candidates in a manner similar to that described by Burke et
al. (2006). Burke et al. validated transit-like features with an ensemble
of features reported by BLS for hundreds of stars observed simultaneously.
Because of the large number of observations in each star's Kepler light curve,
we validate transit-like features in each light curve with the ensemble of
transit- and anti-transit features from only that particular light curve. 

\begin{figure}
\centering
\plotone{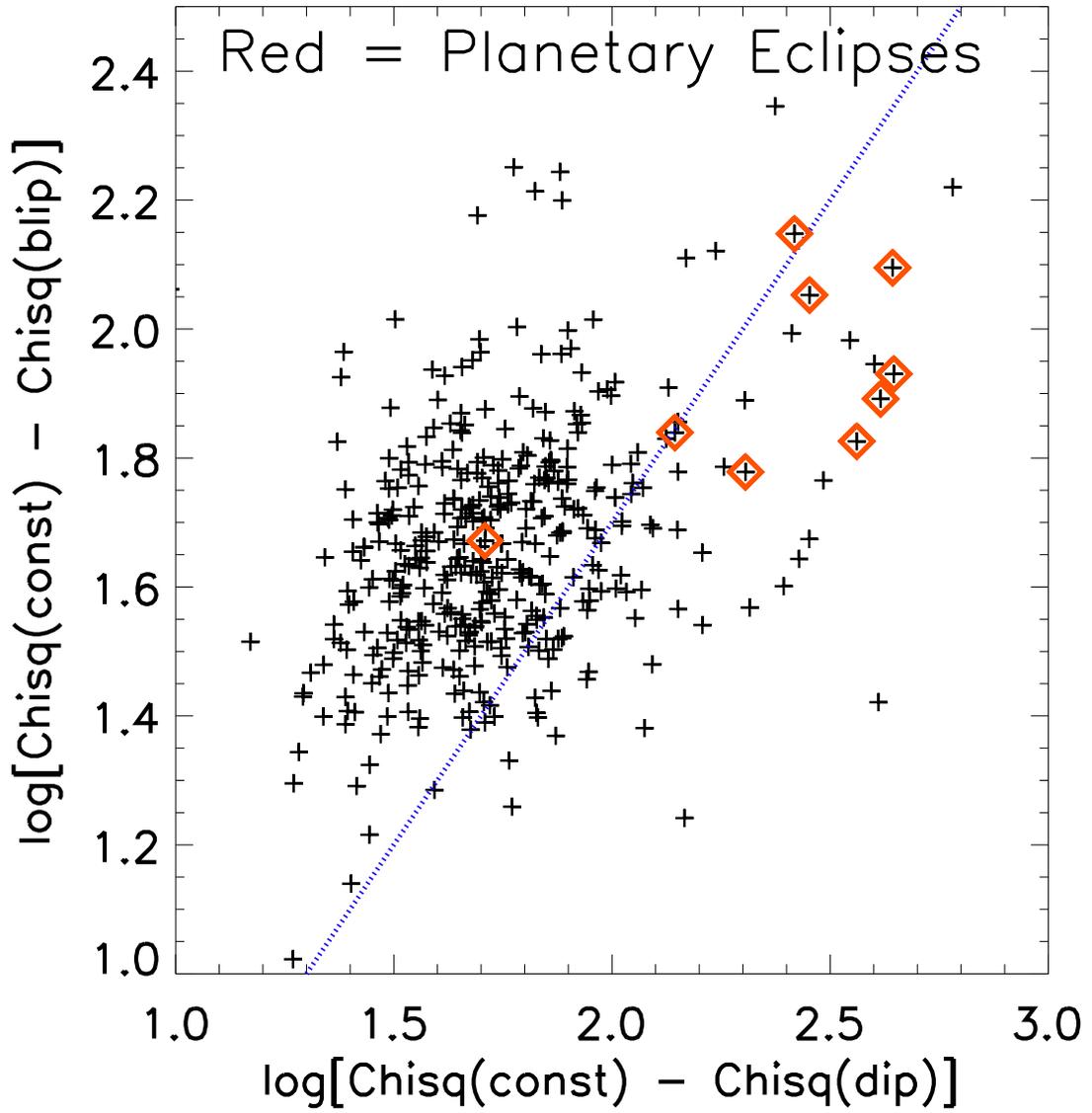}
\caption{
The transit/anti-transit diagram for {\protect \object{KIC 4862625b}}. The red symbols represent the planetary transits and the dotted blue line -- the merit criterion used.
\label{fig:dipblip1}}
\end{figure}

For completeness, we briefly describe Burke's method here. Assuming
dimming features (transits) and brightening features (``anti-transits'')
are due to systematic effects, it is reasonable to expect that there
will not be a strong tendency for such effects to produce dips
versus blips. In other words, typically $\delta_{(\chi^{2}){transit}}$ will
be similar to $\delta_{(\chi^{2}){anti-transit}}$ for each segment.
On the other hand, a highly significant transit signal is
an outlier in a $\delta_{(\chi^{2}){transit}}$ and
$\delta_{(\chi^{2}){antitransit}}$ diagram (Figure
\ref{fig:dipblip1}). The segments where noise dominates (black
crosses) cluster in a cloud of points with similar values for
$\delta_{(\chi^{2}){transit}}$ and $\delta_{(\chi^{2}){anti-transit}}$,
but the segments containing the tertiary transits (red symbols) are
well separated. As seen from the figure (and depending on the merit
criterion) there can be a significant number of outliers,
requiring a human eye to check the segments which triggered
the routine. The number of triggers are highly dependent on the
detrending of the baseline -- quiet stars (like Kepler-16)
have very few outlying points (planetary transits) where smaller
transit signals in more variable binaries (Kepler-47) will be accompanied
by a greater number of false positives. While non-trivial, the number 
of outliers that pass inspection is still much smaller than the total number of 
segments -- typically not more than 10\% of the entire light curve.

\begin{figure}
\centering
\plotone{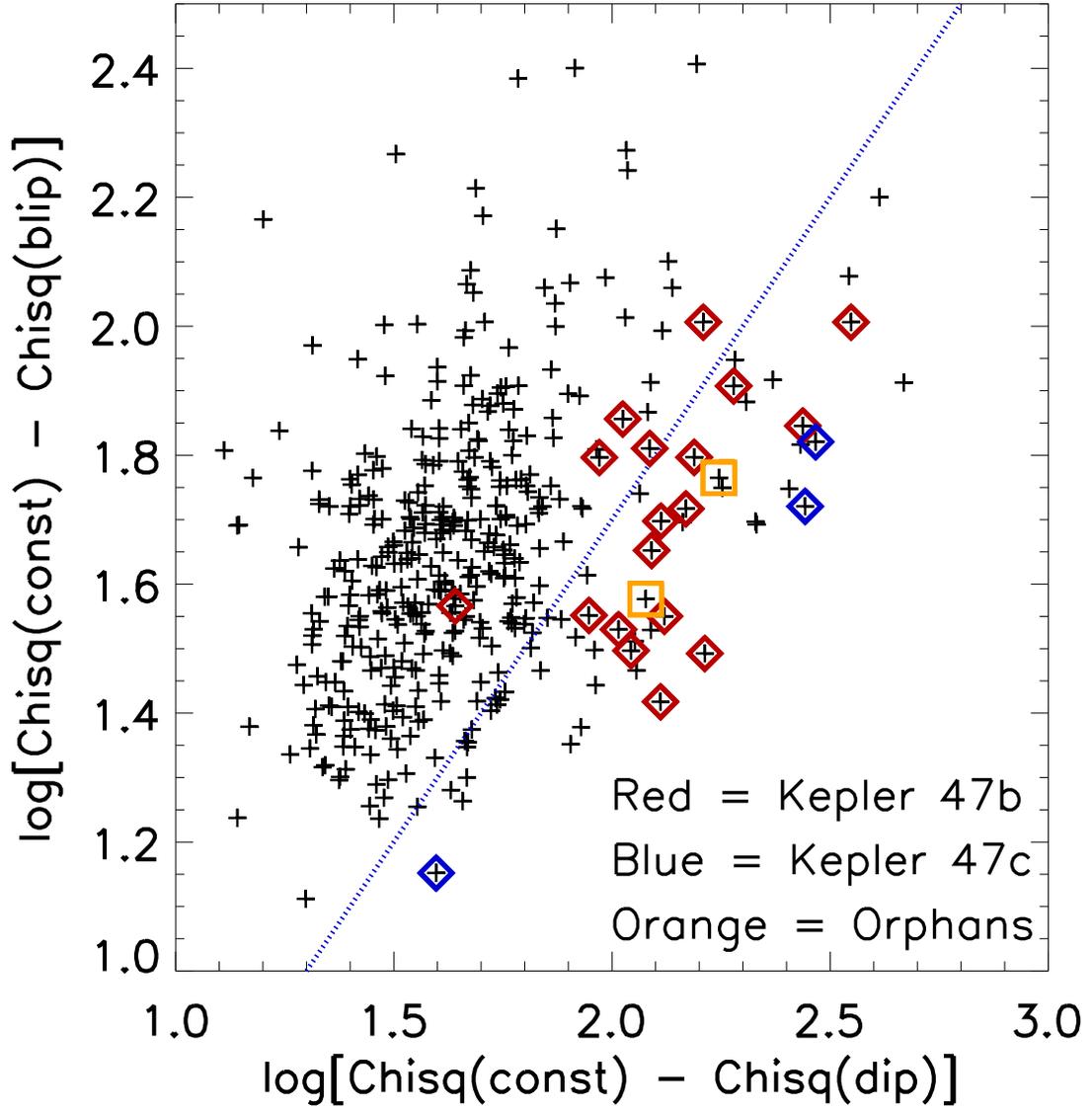}
\caption{
The transit/anti-transit diagram for {\protect \object{KIC 10020423}}, a.k.a. Kepler-47. The different colors represent planets 47b (red) and 47c (blue) and two additional transits (yellow) not associated with either of them.  
\label{fig:dipblip2}}
\end{figure}

\begin{figure}
\centering
\plotone{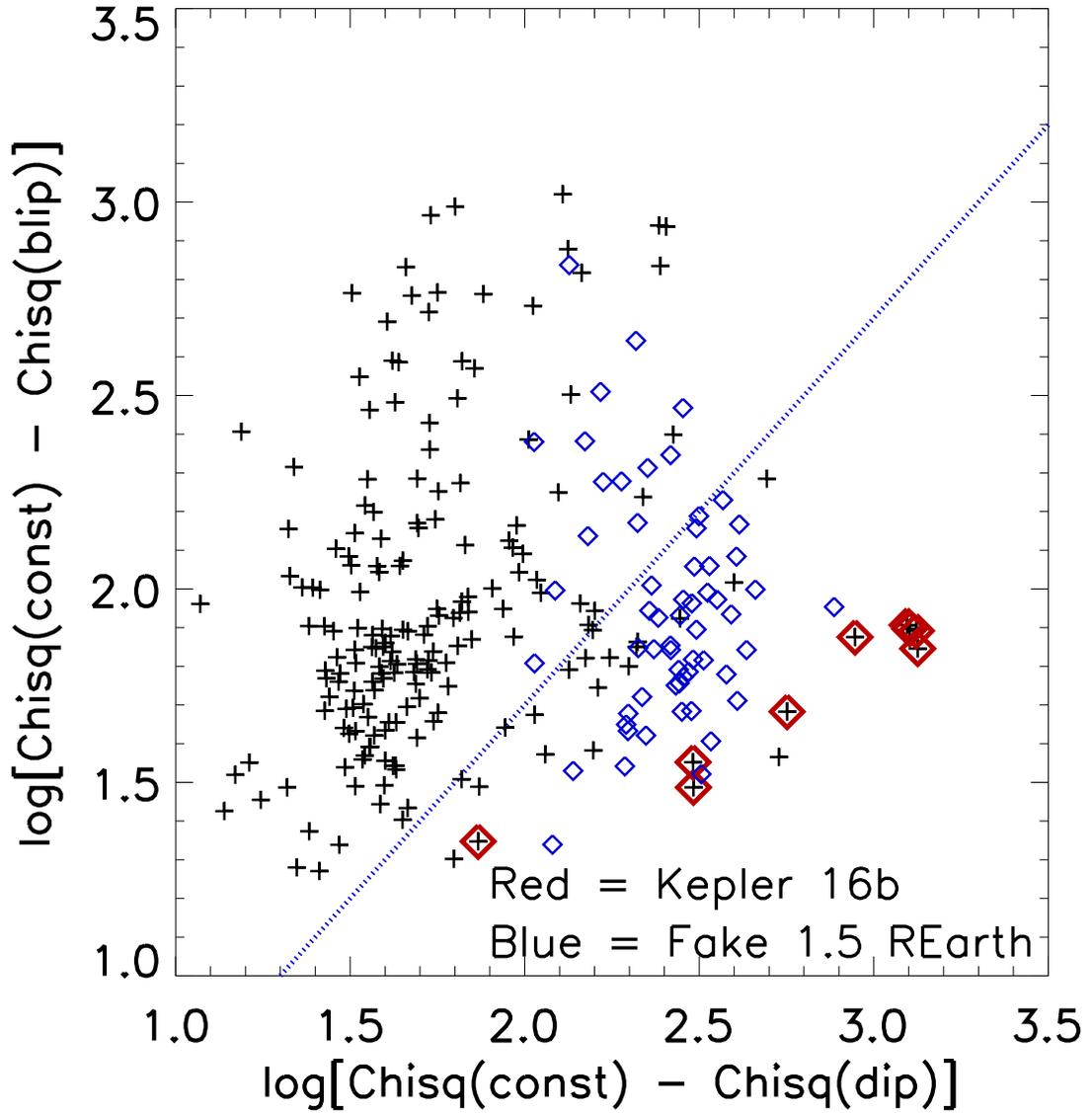}
\caption{
The transit/anti-transit diagram for Kepler-16 with simulated transits of an $1.5R_{Earth}$ planet superposed (blue). The red diamonds represent the transits of Kepler-16b. 
\label{fig:dipblip3}}
\end{figure}

Figures \ref{fig:dipblip1}, \ref{fig:dipblip2} and \ref{fig:dipblip3} show examples from the light
curves for \object{KIC 4862625}, Kepler-47 and Kepler-16 using data from Quarters 1 through 14. \footnote{Earlier versions of the three figures, based on only Quarters 1 to 6, were presented on February 8, 2012 to a committee of Johns Hopkins University as part of the graduate student matriculation procedures for one of us (V.B.K.).}

To evaluate the sensitivity of our method, we insert fake transits
in the light curves and study their recovery rate as a function of
the transit depth. An example for Kepler-16 is shown on Figure
\ref{fig:dipblip3} where the black crosses represents segments with
no fake transits, red symbols are due to Kepler-16b and the blue diamonds
correspond to segments in which fake transits with a depth of 200
ppm ($\sim1.5R_{Earth}$ signal) were superposed. 
The fake transits have variable depths, durations and period to simulate that expected for
a CB planet. Also they all been inserted outside the critical semimajor
axis for stability in the system. 
Not all of the fake transits are recovered: some fall into data gaps, others into noisier parts
of the light curve. However, for the case of Kepler-16,
simulations of super-Earth transits resulted in a 75\%
recovery rate. As seen from Figure \ref{fig:dipblip3}, 
the majority of the inserted fakes occupy a clearly-defined cloud, well-separated from the transit-free segments. 

Their recovery rate is used to adopt the target-specific merit criterion mentioned above, defined in terms of the ratio between $\chi^{2}_{transit}$ and $\chi^{2}_{anti-transit}$ (dotted blue line on Figures \ref{fig:dipblip1}, \ref{fig:dipblip2} and \ref{fig:dipblip3}). Only those segments that fall above an iteratively chosen ratio (2 in the three cases shown here) are inspected visually in the raw data for known systematic features, for centroid shifts and for binary star period commensurability for those systems without well-defined secondary eclipses. As seen from Figure \ref{fig:dipblip1}, one of the planetary transits falls short of the cut criterion with a ratio of 1.01 and another is very close (ratio of 1.1) -- the detrending was not optimal and, due to the long duration of the two transits, significantly dampened them, mistaking them for baseline variability. The transits that do not fulfill the cut criteria for the case of Kepler-47 (Figure \ref{fig:dipblip2}) do so for a different reason -- they are very weak and are difficult to discern from the noise. The sheer number of tertiary events, however, promoted very careful visual examination of the light curves of both targets which ultimately revealed those missed by the automatic procedure as well. This shows that the described method works well for finding not only individual transits but, depending on circumstances, also for super-Earth transits of quiet stars and even for transits with durations as short as only three low-cadence Kepler samples, e.g. the quaternary eclipse of Kepler-16b.

\section{Spectra}
\label{sec:spectra}

\subsection{Apache Point Observatory Observations}
\label{sec:APO}

\object{KIC 4862625} and Kepler-47 were observed with the 3.5~m
telescope at the Apache Point Observatory on four occasions between
April and July, 2012. Coincidentally, the spectroscopic observations
of Kepler-47 by Orosz et al. (2012) and ourselves began within 48
hours of each other. We used the medium dispersion Dual Imaging
Spectrograph (DIS) in its highest resolution spectroscopic mode.
The pair of B1200/R1200 gratings in combination with the 1.5$\arcsec$
slit delivers a spectrum with resolution R$\sim$3000 and covering simultaneously two
1200\AA\ windows centered on 4500\AA and 6500\AA, respectively. 
The slit was oriented along the parallactic angle. 
Each night we obtained one or two  exposures of each target supplemented by exposures of
spectrophotometric and radial velocity standards.
Nightly we also obtained several HeNeAr lamp spectra for wavelength calibration, and
we used telluric lines in the observed spectra to correct for offsets due to flexure or other instrumental effects. 
Conditions of all four nights were not strictly photometric. 
Each target was observed
for $\sim$900 seconds per night, yielding a peak signal to noise ratio
in the continuum of 20 to 30 per resolution element.

The data reduction included bias and flat-field correction, aperture extraction, wavelength and flux calibration. 
We compared the APO long-slit spectrum of \object{KIC 4862625} with a library of
stellar spectra (Pickles 1998).  For the R$\sim$3000 APO spectra,
the FWHM $\sim$45 km s$^{-1}$ broadening evident in the SOPHIE spectra (Section \ref{sec:sophie})
is unresolved. We estimate from the shape of the continuum and strengths of particular
spectral lines that the best match is spectral type F8 IV. 
The subgiant classification is consistent with the density of star A determined later from the
light curve and the mass function. The Kepler Input Catalog (KIC) tends to over estimate log(g) for subgiants, especially
those hotter than about 5400 K, which can lead to underestimates of their radii in the KIC, typically by factors of 1.5 to 2 (Brown et al. 2011).
For \object{KIC 4862625}, the KIC radius estimate is 0.806 R$_\odot$, indeed approximately half the size that we estimate in this work.

\subsection{SOPHIE observations and data reduction}
\label{sec:sophie}

The two targets were observed at the end of summer 2012 with the SOPHIE spectrograph at the 
1.93-m telescope of Haute-Provence Observatory, France. The goal was to detect the reflex 
motion of the primary stars due to their secondary components through radial velocity 
variations. 
SOPHIE (Bouchy et al.~2009) is a fiber-fed, cross-dispersed, environmentally stabilized 
echelle spectrograph dedicated to high-precision radial velocity measurements. 
For such binary systems the amplitudes of variation are expected to be of the 
order of a few to a few tens  km\,s$^{-1}$, which is well in SOPHIE capabilities despite the 
faintness of the targets.
The data were secured in High-Efficiency mode (resolution power $R=40\,000$) and slow 
read-out mode of the detector. In order to reach signal-to-noise ratio per pixel of the order of 10 
at 550~nm, exposure times ranged between 1200 and 2000~sec for Kepler-47, and between 
500 and 900~sec for \object{KIC 4862625}.

The spectra were extracted from the detector images with the SOPHIE pipeline, 
which includes localization of the spectral orders on the 2D-images, optimal order extraction, 
cosmic-ray rejection, wavelength calibration and corrections of flat-field. 
Then we performed a cross-correlation of the extracted spectra 
with a G2-type numerical mask including more than 3500 lines, and 
finally measures the radial velocities from Gaussian fits of the cross-correlation 
functions (CCFs) and the associated photon-noise errors,  following the method described 
by  Baranne et al.~(1996) and Pepe et al.~(2002).
For Kepler-47 and \object{KIC 4862625} respectively, the full widths at half 
maximum of those Gaussians are $12 \pm 1$~km\,s$^{-1}$ and 
$15 \pm 2$~km\,s$^{-1}$, and their contrasts are
$17 \pm 4$\,\%\  and $4 \pm 2$\,\%\  of the continua.
One of the observations of \object{KIC 4862625} was made at the twilight: the pollution due to the bright 
sky background was corrected thanks to the reference fiber pointed on the sky 
(e.g. H\'ebrard et al.~2008). Other exposures
were not polluted by sky background nor Moon light.
In Table~\ref{tab:RadVel} the SOPHIE radial velocities are absolute in
barycentric reference, whereas the APO radial velocities are absolute for \object{KIC 4862625} and relative for Kepler-47.

Simultaneously with the publication of our results, Schwamb et al.~(2012) published their study of KIC4862625. 
The main difference in the two analyses is that we assumed there is only one star in the 
SOPHIE aperture, whereas Schwamb et al.~(2012) have shown that there is a 
significant contaminant at 0.7~\arcsec from the primary star, that contaminant being 
itself a binary. The contaminant is well within the 3~\arcsec SOPHIE aperture. Whereas 
it is clearly detected in the broadening function of the HIRES spectra of Schwamb et al.~(2012)
(their Figure 7), it is not as clear in the SOPHIE CCFs. 

With that in mind we reinspected the SOPHIE CCFs. 
By letting a second peak free to vary in each of the eight SOPHIE spectra, we did not 
significantly detect it. However, by assuming that second peak is not varying with 
time (the variations seen with HIRES are small), it is detected in the SOPHIE 
spectra at a radial velocity of $-22.7\pm0.7$~km\,s$^{-1}$, 
FWHM~$=8.9^{+2.4}_{-1.6}$~km\,s$^{-1}$ and a contrast of 
$=0.53^{+0.14}_{-0.17}$\,\%\ of the continua.
So there is a 3.8-$\sigma$ detection of the second peak on the total data of the eight SOPHIE spectra.
We revised the SOPHIE radial velocity measurements of KIC4862625 taking into 
account for that additional peak in the CCFs. The resulting radial velocities differ by at most 2$\sigma$ from the radial velocities obtained without taking into account the second peak. The final radial velocities of both stars are reported in Table~\ref{tab:RadVel}. 

Radial velocity variations in phase with the Kepler ephemeris are clearly detected. Section \ref{sec:orbit}
addresses the orbital parameters derived from the radial velocities in combination with the light curve.
Section \ref{sec:spectralanalysis} addresses atmospheric parameters of star A derived from the spectra.
\section{Data Analysis}
\label{sec:data}

\subsection{Eclipsing Binary Light Curve}
\label{sec:eblc}

\begin{figure}
\centering
\plotone{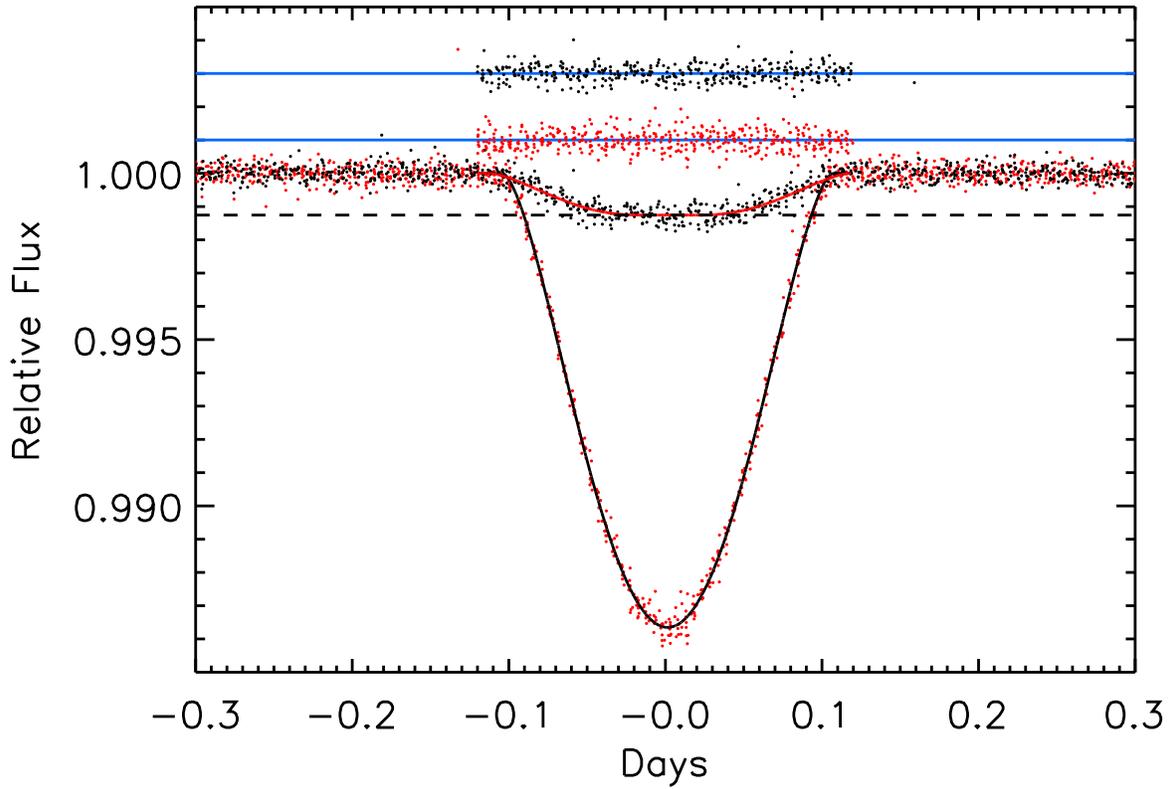}
\caption{
Kepler light curves of {\protect \object{KIC 4862625}}. The normalized and detrended flux of the primary and secondary eclipses are shown
with respect to the orbital phase of the EB, along with the binary star model. The secondary
eclipse data have been centered at zero phase for comparison with the primary eclipse data. The secondary
eclipse data and the residuals (above) have been offset vertically for clarity.
\label{fig:lc}}
\end{figure}

The light curve of the EB star constrains the {\it relative} orbit of the two stars:
the orbital period $P$, the center times of primary eclipse $T_t$ and secondary eclipse $T_o$,
the semi-major axis of the relative orbit in units of the radius of star A, $a/r_A$, 
and the orbital inclination, $i$. Also from the light curve, we derive the relative radii of the
two stars, $k = r_B / r_A$, and the fraction of the flux in the Kepler bandpass\footnote{http://keplergo.arc.nasa.gov/CalibrationResponse.shtml} emitted
by star B, $f_B$. Nominally, we assume zero ``third light,'' so $f_A = 1 - f_B$; we examine that assumption later in this section.
We adopt the period $P = 20.000214$ d from Pr{\v s}a et al. (2010), although the trend
in the eclipse timing variations suggest a period $\sim$2 seconds longer.
The free parameters of the fit to the light curve are $T_t, T_o, a/r_A, p, i, {\rm and} f_B$.
We compute the fraction of light blocked by one star by the other using computer
code of Mandel \& Agol (2002). Because the latter code was designed for planetary transits
and models the nearer body as an entirely dark and opaque circular disk, we account for
the light from the nearer star appropriately by superposition. The phase-folded light curve is shown on Fig.~\ref{fig:lc}

We estimate the limb darkening parameters $(u_{1,A},u_{2,A}) = (0.243,0.371)$ and $(u_{1,B},u_{2,B}) = (0.112,0.350)$ for stars A and B respectively, by interpolation of tables of R-band theoretical quadratic limb-darkening coefficients (Claret 2000) as appropriate for their effective temperatures, gravities, and metalicities (Table \ref{tab:parameters}). For star A, those are estimated from spectroscopy (Section \ref{sec:spectralanalysis}). For star B, we estimate its effective temperature $T_{eff} = 3390 \pm 50$ K, by interpolating a grid of model atmosphere spectra 
(Hauschildt et al. 1999\footnote{Machine-readable tables are available at http://svo.cab.inta-csic.es/theory/db2vo/index.php .}) to match the depth of the light curve at secondary eclipse; we integrate the spectra over the Kepler bandpass and compare to that of star A, accounting also for the relative solid angles, $k^2$.
We estimate star B's gravity log(g) = 4.94 in cgs units from its mass $\vMstarB \pm \eMstarB\ M_\odot$ and
radius $\vRstarB\ \pm \eRstarB\ R_\odot$.
We assume star B's metalicity is equal to that of star A.

We derive the quantity $e (\cos\omega) / \sqrt{1-e^2}$ from the phase
of the center of the secondary eclipse relative to the center of primary eclipse.
For the latter constraint, we used the analytic approximation for the case of
inclination $i = 90\arcdeg$ (Hilditch 2001, Eq. 5.67), and verified that the difference between that
approximation and the numerical estimate for $i > 87\arcdeg$ is negligible (Hilditch Eq. 5.63).
Similarly, the quantity
$e sin \omega$ equals the ratio of the difference to the sum of the durations of secondary and
primary eclipses (Hilditch Eq. 5.69). However, because the eclipse durations are much less precisely
measured than the centers, we do not explictly constrain $e sin \omega$, although it is
weakly constrained implicitly in fitting the light curve. Instead, $\omega$ is measured
better using the radial velocities (Section \ref{sec:orbit}).\footnote{We have adopted the equations
and viewing geometry of Hilditch's textbook, Figure 2.5. Apparently, prior publications of CB planets have adopted
the opposite viewing geometry, e.g. Fig. 7 of Murray \& Correia (2010). This
affects $\omega$ by 180\arcdeg.}

Given the above constraints from the light curve, fitting the radial velocities 
depends on only three astrophysical free parameters: the systemic velocity $\gamma$,
the velocity semi-amplitude $k_1$, and the longitude of periastron $\omega$. In our analysis,
once $\omega$ is measured using the radial velocities, the eccentricity $e$ and
the time of periastron passage $t_{peri}$ are analytically constrained from the
light curve (Hilditch Eqs. 4.10 and 5.67). Similarly, the uncertainties in the parameters
$e$ and $t_{peri}$ flow down from the uncertainty in $\omega$ determined from the
radial velocities.  

The planetary transits of star B, if they occur, are undetectable in the Kepler photometry.
Because star B contributes only $f_B = 0.00124$ of the flux of the system, and its relative solid angle
$k^2 = 0.0426$, the mean surface brightness of star B is 0.029 that of star A, in the Kepler
bandpass. Given that the
planet blocks $\sim$0.1\% of the system's light when transiting star A, we predict only a 30 ppm planetary
transit of star B, i.e. undetectable with Kepler's per-cadence RMS   noise of 222 ppm.
Because CB transit durations can be no longer than
half of the orbital period of the EB (Schneider \& Chevreton 1990), or 10 days in this case,
and even with an idealized 10-day upper-limit on the duration, the 30-ppm transit depth would
correspond to ${\sim}3\sigma$, and typically not even that, due to long-term intrinsic variations in the system's total light.

In the above analysis, we have assumed that the Kepler photometer records the sum of the light from the two stars.
star A and star B, and nothing more, i.e. zero ``third light''. In our spectroscopic observing, we took care to inspect
images from the acquisition cameras, and noted no stars of any significance within the range $1-2$\arcsec\ of
\object{KIC 4862625}; 2MASS images support this, also. Because poor weather thwarted an attempt at adaptive optics imaging of the environs of \object{KIC 4862625},
we were unable to inspect within the $\sim$1\arcsec\ seeing limit. In general, however, imaging can never prove there is zero third light, because any system could be a hierarchical triple star. 

We investigated the effects of assuming a given third-light fraction, $f_C$ = 0, 0.1, and 0.2, of the total light. For $f_C = 0$ the parameters are $k = r_B/r_A = 0.196$, $a/r_A = 21.6$, and $i = 87.53$\arcdeg, for $f_C = 0.1$ they are $k = r_B/r_A = 0.201$, $a/r_A = 22.08$, and $i = 87.59$\arcdeg, and for $f_C = 0.2$ the parameters are $k = r_B/r_A = 0.227$, $a/r_A = 23.56$, and $i = 87.60$\arcdeg. Thus, compared to the nominal $f_C = 0$ solution, non-zero third light implies larger stellar densities and a larger star B relative to star A. As pointed out by Schwamb et al.~(2012), submitted simultaneously with our results, there indeed is a ``third light'' contamination to the light curve of KIC 4862625 in the form of another binary system $0.7\arcsec$ away, with a flux contribution of $\sim 10\%$ in the {\it Kepler} bandpass. Thus, throughout the paper we use the solution for $f_C = 0.1$, listed in Table \ref{tab:parameters}.

\begin{figure}
\centering
\plotone{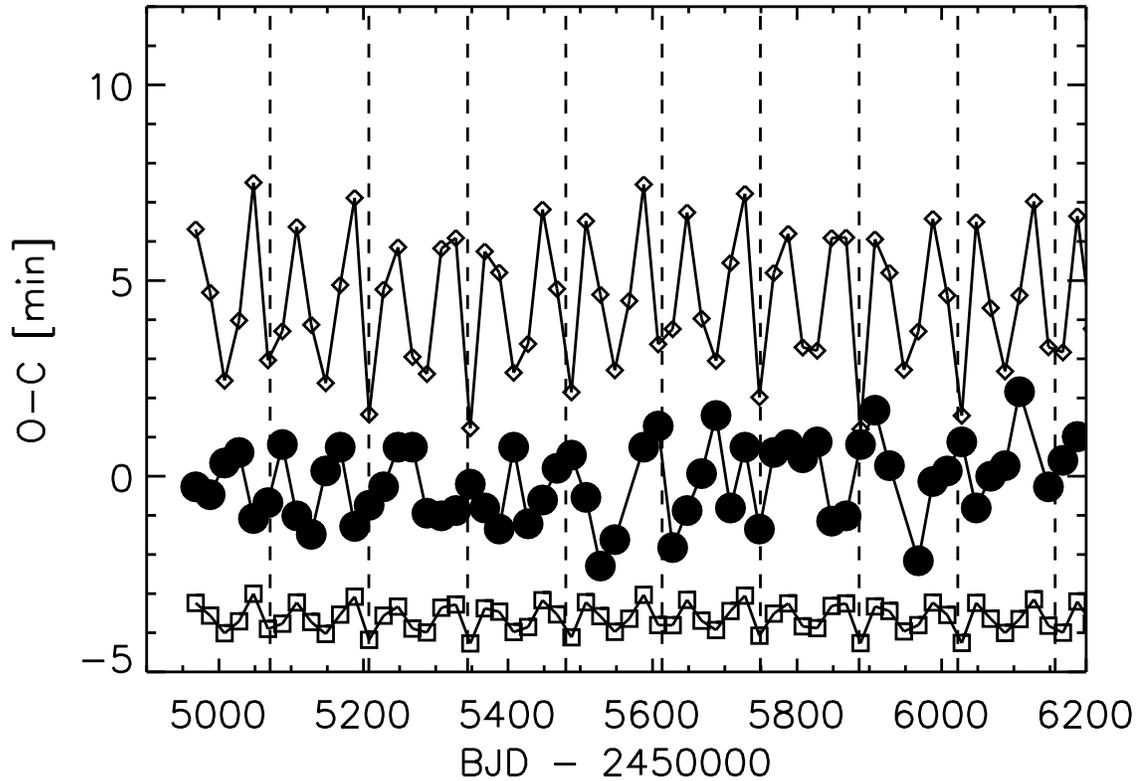}
\caption{
Eclipse timing variations of the primary eclipses of {\protect \object{KIC 4862625}} (filled circles). The observed times of each primary eclipse minus the calculated times for a linear ephemeris are shown versus time (an ``O-C'' diagram). The nine planetary transit events are indicated by vertical dashed lines. A single primary eclipse is missing from the sequence at BJD = 2455567.8 and another, at BJD = 2455947.8, is incomplete. For comparison, two simulated ``O-C'' variations are also shown, one for a 1 $\Mjup$ (squares) and the other for a 5 $\Mjup$ (diamonds) respectively, and shifted vertically for clarity. Evidently, the circumbinary body's mass is less than 5 $\Mjup$. 
\label{fig:etv}}
\end{figure}

To estimate eclipse time variations, ETVs (Figure \ref{fig:etv}), we empirically determine the best-fitting center
time of each primary eclipse individually by minimizing $\chi^2$, while
adjusting only the center time of the model curve. The model is the Mandel \& Agol (2002)
curve that best-fits the ensemble of primary eclipses, although simpler, non-physical models
such as Gaussians or U-shaped curves produce very similar ETVs. We evaluated the statistical
significance of the measured ETVs by simulation of model eclipses superposed on the 
detrended Kepler Light curve. We fit the simulated eclipses to produce simulated ETVs, which we found to have
slightly smaller amplitude than the actual ETVs. We conclude that the measured ETVs
may have a contribution from the mass of the gas-giant planet but are essentially consistent with noise (Figure \ref{fig:etv}). Also, we simulated ETVs for our best-fit model (Section \ref{sec:results}) for tertiary masses of 1, 5 and 50 $\Mjup$ and compared them to the measured values; the latter fall between those for 1 and 5 $\Mjup$, securing the planetary nature of the circumbinary body.

\subsection{Spectral analysis}
\label{sec:spectralanalysis}

The SOPHIE spectra of \object{KIC 4862625} without background pollution were co-added for spectral analysis.
The H$\alpha$ and H$\beta$ lines were used to determine the effective temperature 
$T_{\rm eff} = 6200 \pm 150$\,K. 
This estimate was made on each line independently in order to check  the consistency of the results.
The spectrum lacks prominent spectral features due to its broad lines combined with the low signal-to-noise ratio. 
This prevents us from carrying out a detailed spectroscopic analysis.
We could not derive accurate estimate of the surface gravity from the Mg\,{\sc i} triplet and the Na\,{\sc i} doublet.
The estimation from these lines is $\log g \simeq 4.0 \pm 0.2$; it is a typical value for main sequence and subgiant stars in that 
$T_{\rm eff}$ range. 
We do not find any evidence of Lithium in the spectrum nor any sign of chromospheric activity 
in the Ca\,{\sc ii} H and K lines, but the H$\alpha$ core shows some variable emission features.
From the width of the CCF we derived  $v \sin i_* = 31\pm2\,$km\,s$^{-1}$ and [Fe/H]\,$\simeq -0.15$.

Using these values of $T_{\rm eff}$, $\log g$, and  [Fe/H], we estimated the mass and radius of the star 
by comparison with a grid of STAREVOL stellar evolution models 
(A.~Palacios, priv. com.; Lagarde et al.~2012). 
We generated a series of Gaussian random 
realizations of $T_{\rm eff}$, [Fe/H] and $\log g$, and for each realization we determined the best evolutionary 
track using a $\chi^2$ minimization on these three parameters. 
We found $M_{\star} = 1.23 \pm 0.20\, \mathrm{M}_{\odot}$, 
$R_{\star}  =  1.70\pm 0.25\, \mathrm{R}_{\odot}$, and an isochronal age of 
$2.6 ^{+3.6}_{-0.3}$\,Gyr. 

\subsection{Orbital solution of the binaries}
\label{sec:orbit}

For each targets, the APO and SOPHIE radial velocities were fit simultaneously with a Keplerian model. 
The fits are mainly constrained by the SOPHIE data, which are more numerous and accurate. The APO radial 
velocities are much less accurate, but agree with the orbital solution derived from the SOPHIE data. 

In addition to the measured radial velocities, the fits take into account the three constraints derived from 
the Kepler photometry: specifically the orbital period $P$, and the mid-times of transit and occultation, $T_t$ and $T_o$.
The latter parameters strongly constrain $e (\cos\omega) / \sqrt{1-e^2}$ (Section \ref{sec:eblc}).
The radial velocities confirm the orbital eccentricities and allow $e$ and $\omega$ to be measured individually.
In comparison to the radial velocity uncertainties, the uncertainties in the three 
parameters derived from photometry have negligible effects on the final uncertainties of the derived 
orbit parameters.

\begin{figure}
\centering
\plottwo{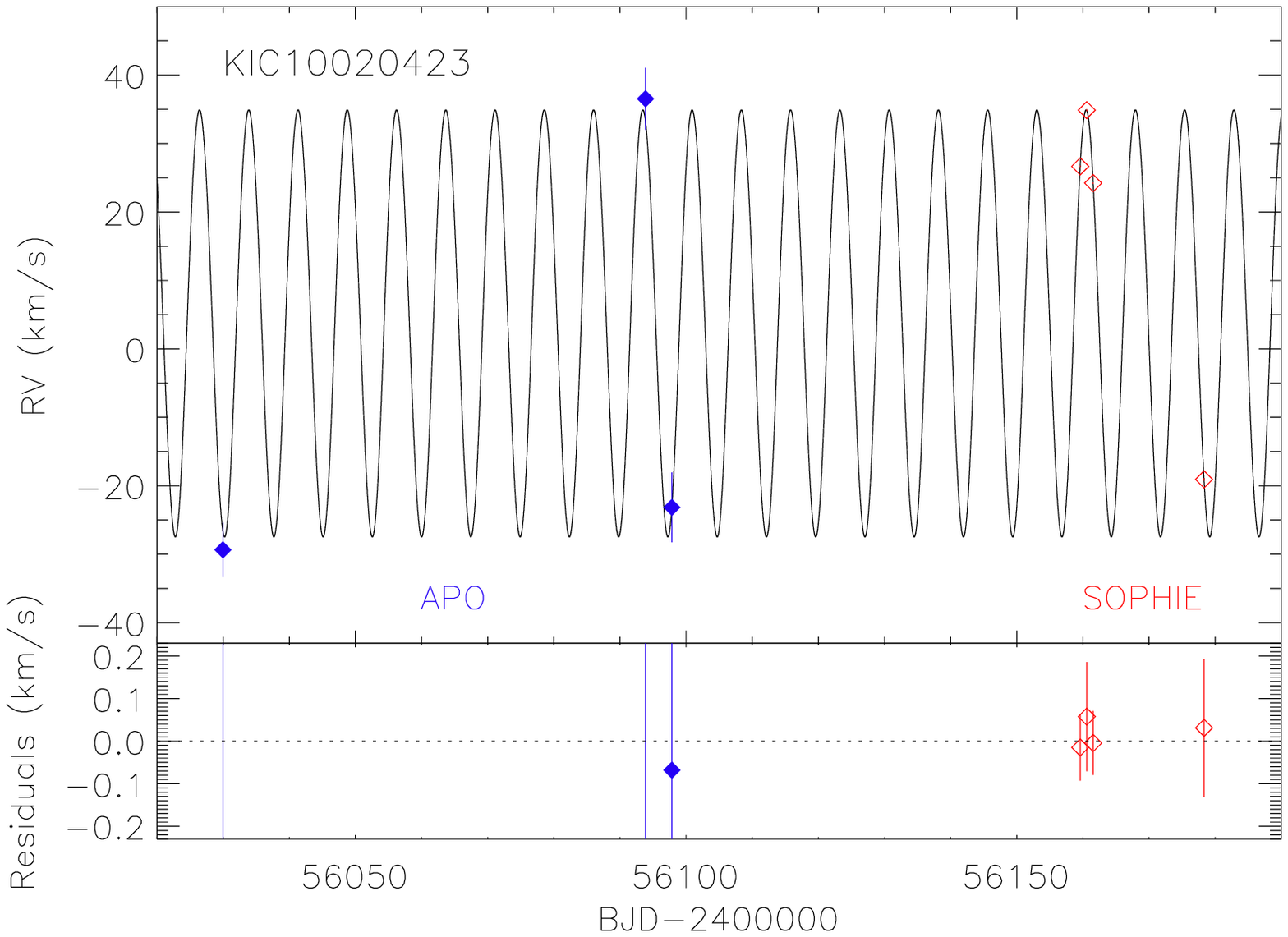}{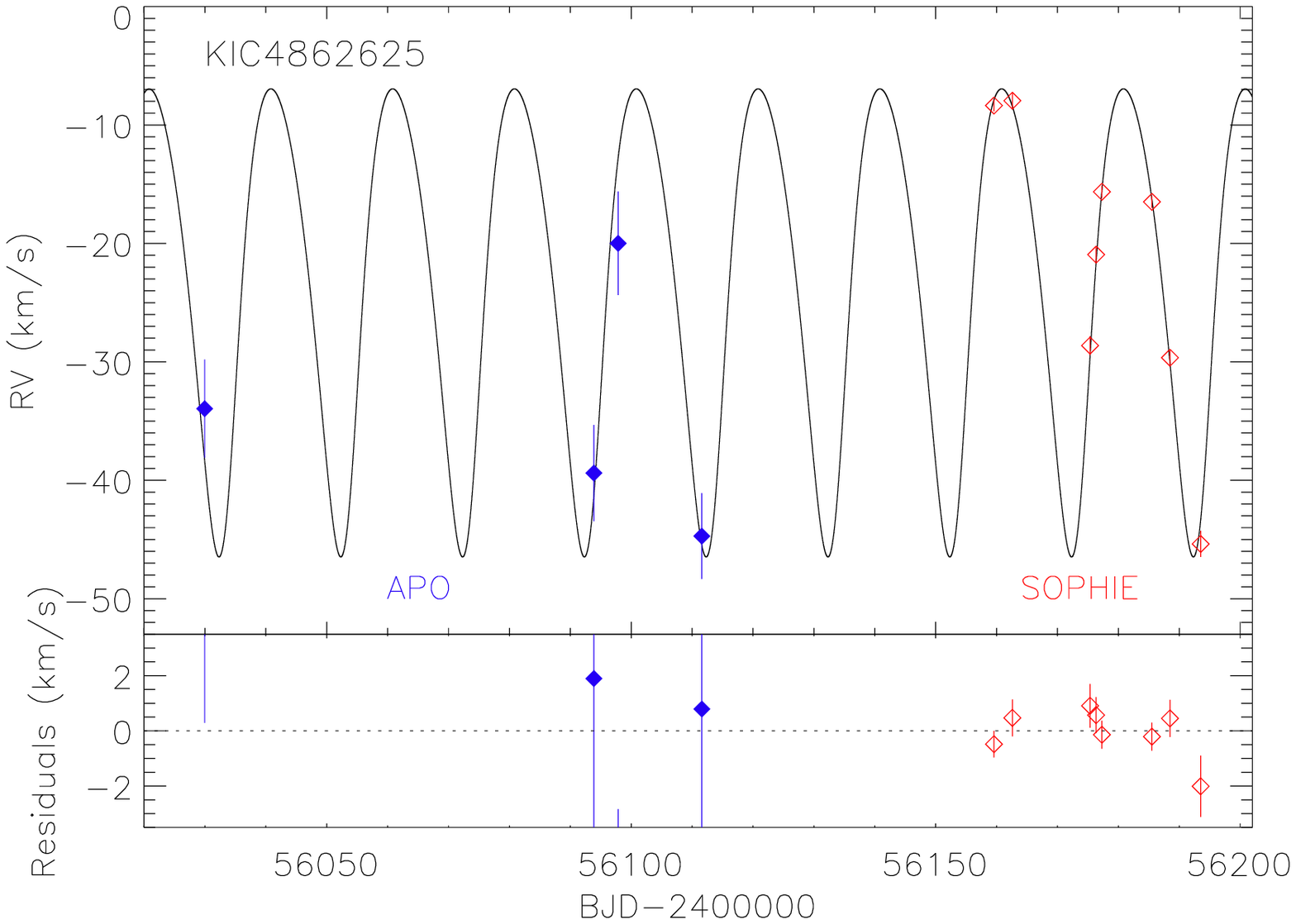}
\plottwo{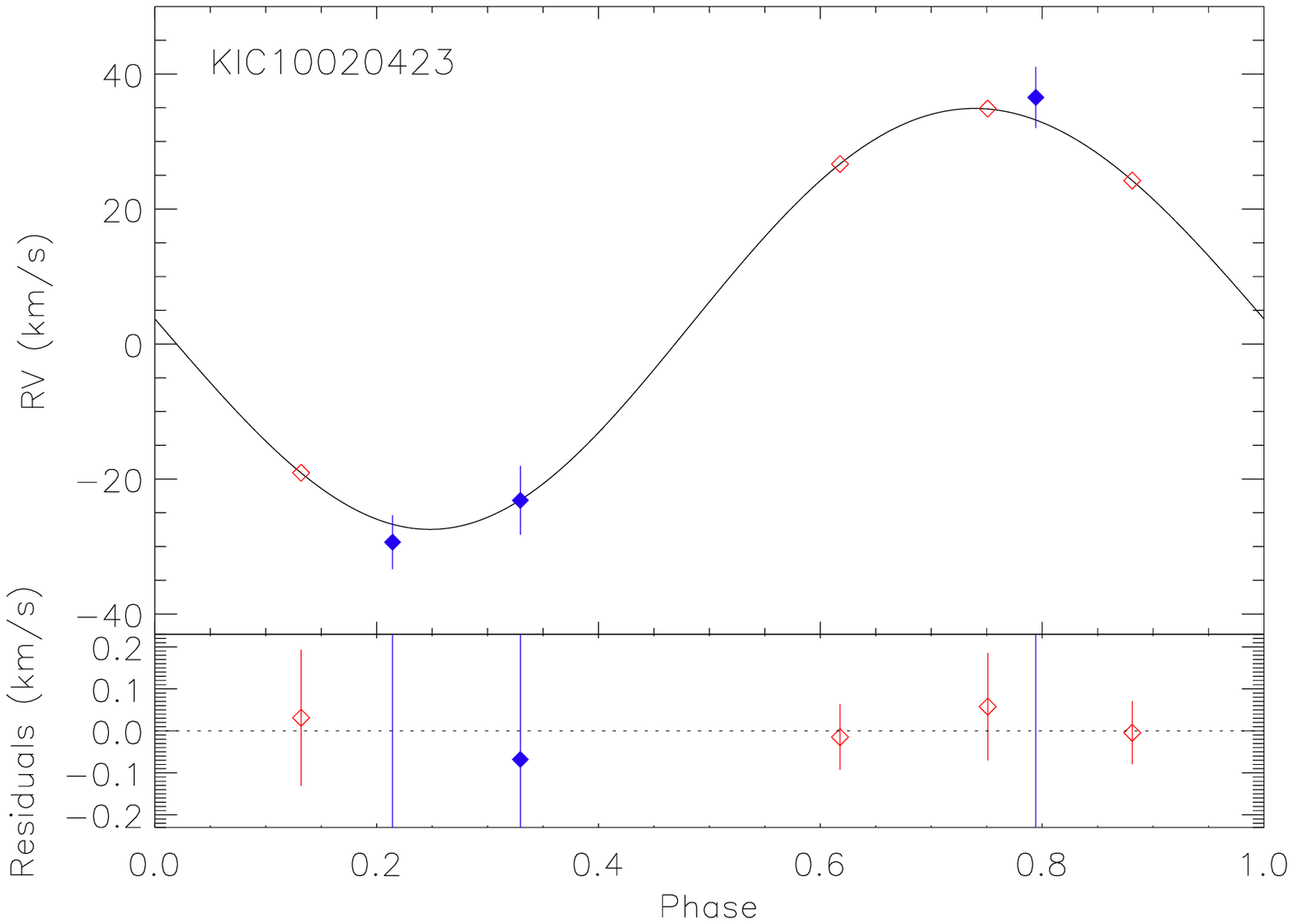}{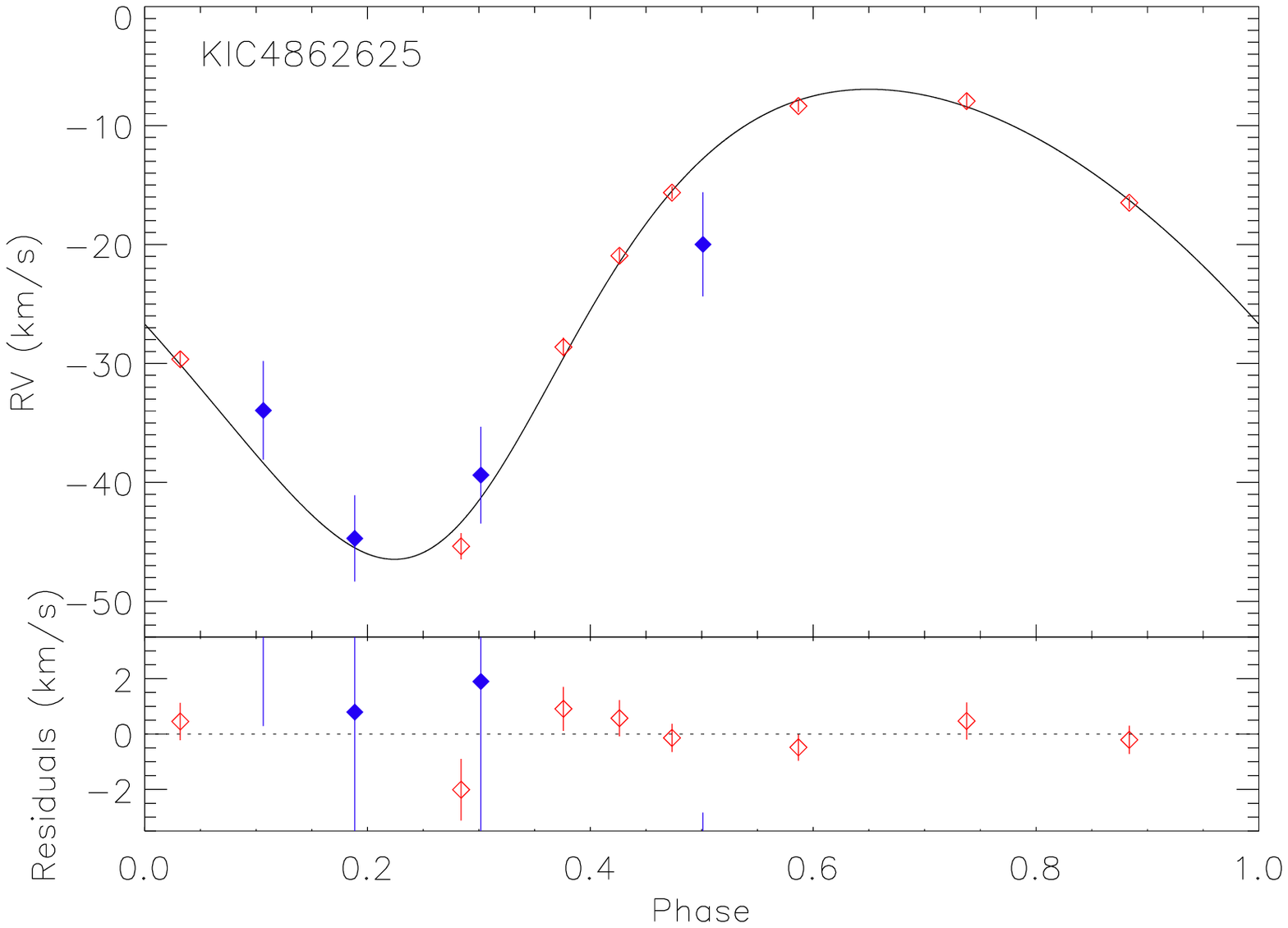}
\caption{Radial velocity measurements of {\protect \object{KIC 10020423}} (Kepler-47, upper and lower left) and {\protect \object{KIC 4862625}} (upper and lower right)
with 1-$\sigma$ error bars as a function of time (upper) or orbital phase (lower) together with their 
Keplerian fit and residuals of the fit. 
The data are from APO (blue, filled diamonds) and SOPHIE (red, empty diamonds).
\label{fig:rv_orbits}}
\end{figure}

The fits to the radial velocities were made using the Levenberg-Marquardt method, and the confidence intervals around the best solutions were determined both from $\chi^2$  variations and Monte Carlo simulations. 
The histograms of the obtained parameters have a single peak and nice Gaussian-like appearance. 
The derived values and uncertainties are reported in Table~\ref{tab:parameters}; 
the best fits are over-plotted with the data in Fig. \ref{fig:rv_orbits}.

The semi-amplitude of the radial velocity variations of 
\object{KIC 4862625} is $K=19.76\pm0.73$ \,km\,s$^{-1}$. This is 1.9\,$\sigma$ larger than the $K$-value 
we obtained by assuming only one component in the SOPHIE CCF instead of two (see above in Section \ref{sec:spectralanalysis}). This gives an order of magnitude of the small effect induced by the blend. Our result is smaller but agrees at 1.9\,$\sigma$ with the $K$-value reported by Schwamb et al.~(2012).
With an adopted primary mass $m_A = \vMstarA \pm \eMstarA \, \mathrm{M}_{\odot}$, this translates in a secondary mass of $m_B = \vMstarB \pm \eMstarB\ \, \mathrm{M}_{\odot}$.
Because \object{KIC 4862625} is single-lined spectroscopic binary, star B's mass depends on star A's and their uncertainties are similarly coupled.
For \object{KIC 4862625} the dispersion of the residuals of the fits are 500\,m\,s$^{-1}$ and 4.3\,km\,s$^{-1}$ for SOPHIE and APO data, respectively. 
The dispersion is similar or even smaller than the expected uncertainties on the 
measured radial velocities. Uncertainties on the SOPHIE radial velocities might be slightly overestimated.
We did not reduce them however in order to be conservative.
We cannot detect any significant drift in addition to the reflex motion due to the binaries. For \object{KIC 4862625}
we estimate an upper limit $\pm10\,$km\,s$^{-1}$\,yr$^{-1}$ for any additional drift.

Our derived parameters for Kepler-47 agree with those derived by Orosz et al.~(2012).
The semi-amplitude of the radial velocity variations of 
Kepler-47 is $K=31.18\pm0.12$ \,km\,s$^{-1}$. With an adopted primary mass from Orosz et al. (2012),
$m_A = 1.04 \pm 0.06 \, \mathrm{M}_{\odot}$, this translates in a secondary mass of $m_B = 0.357 \pm 0.013 \, \mathrm{M}_{\odot}$.
Because there are so few RV measurements for Kepler-47, their dispersion is less than the expected measurement uncertainty.
For Kepler-47 the dispersion of the residuals of the fits are 25\,m\,s$^{-1}$ 
and 2.6\,km\,s$^{-1}$ for SOPHIE and APO data, respectively. 

\subsection{Doppler Boosting}
\label{sec:boosting}

\begin{figure}
\centering
\plotone{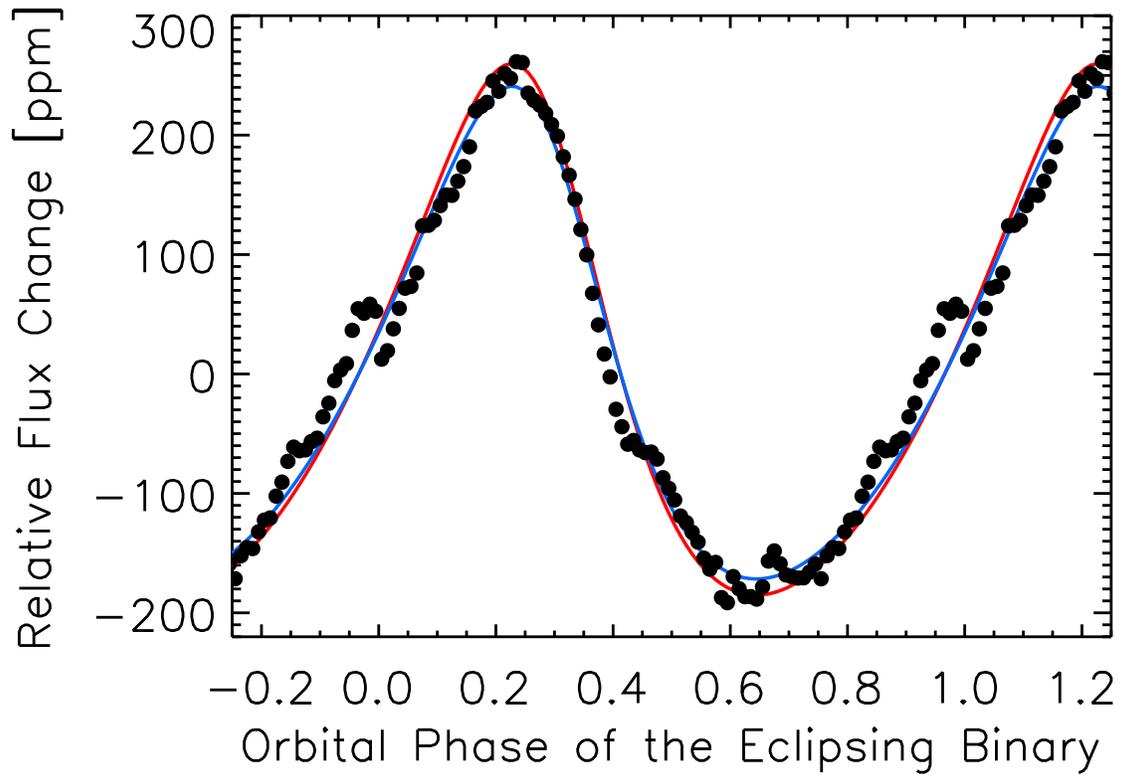}
\caption{
Doppler boosting of {\protect \object{KIC 4862625}}. 
The relative, normalized flux change (see text) is plotted with respect to the EB's
orbital phase. The data binned in 0.01 intervals of phase (filled circles) approximately
match the flux change estimated from Doppler boosting (red curve) based upon the spectrum of
the primary star and its spectroscopically-determined radial velocity curve.
The amplitude of the best-fitting curve (blue) is $93 \pm 1.7$\% that of the estimate (red).
\label{fig:boosting}}
\end{figure}

The oscillating radial velocity of the primary star of \object{KIC 4862625} is apparent in the 
Kepler photometry. Due to Doppler boosting, when the star is moving toward the Earth, its observed flux increases and 
when the star is moving away, its observed flux decreases. Figure \ref{fig:boosting} illustrates
the modulation in observed flux as a function of the orbital phase of the EB star.
To reduce the effect of the rotational modulation on the light curve, we used the mean flux level 
estimated at each point from the $\sin$ wave fit at each point (Section \ref{sec:detrending}). 
We grouped the results in 100 uniformly spaced bins in orbital phase. Hence, each point in
Figure \ref{fig:boosting} represents the median of $\sim 350$ Kepler measurements. With an RMS
of 222 ppm per original Kepler observation, each median would have a formal uncertainty $\sqrt{350}$ less,
or 12 ppm if there were no trends in the light curve. However, with the trends, the observed RMS   
deviation of the medians with respect to the best-fitting boosting curve is 19 ppm.

Due to Doppler boosting, the ratio of observed flux $F_\lambda$ to emitted flux $F_{0,\lambda}$ is 
\begin{equation}
{ {F_{\lambda}} \over {F_{0,\lambda}} } = 1 - B {v_r \over c},
\label{eq:boost}
\end{equation}
where $v_r$ is the stellar radial velocity, c is the speed of light, and the Doppler boosting factor
$ B = 5 + d ln F_\lambda / d ln \lambda $ (Bloemen et al. 2010; Loeb \& Gaudi 2003). 
For a T=6150 K blackbody approximation to the star A's spectrum, and a monochromatic approximation
to the Kepler bandpass of $\lambda = $ 600 nm, the boosting factor $B_{BB} = 3.99$ 
(Loeb \& Gaudi 2003, Eqs. 2 and 3). At a finer level of approximation, using a template spectrum for
an F8 IV star (Pickles 1998) and the Kepler bandpass, we estimate a photon
weighted bandpass-integrated boosting factor $B_{F8 IV} = 3.73$ (Bloemen et al. 2010, Eq. 3). Both estimates
neglect reddening, but its effect is very small for interstellar reddening typical of Kepler stars
(Bloemen et al. 2010). Figure \ref{fig:boosting} compares the Doppler boosting effect estimated 
with $B = B_{F8 IV}$, to the Kepler photometry of \object{KIC 4862625}. The boosting factor that best fits the
Kepler photometry and the spectroscopic radial velocity curve is $B = 3.46\pm 0.065$, i.e.
or $93 \pm 1.7$\% of the analytic estimate with $B = B_{F8 IV}$. The boosted flux from star B is
out of phase with that of star A, but for simplicity of this analysis we have neglected the tiny
contribution from star B.

The capability to measure the radial velocity of an EB star using Kepler data alone
could be useful and convenient. In principle, CB systems could be ``solved'' without
spectroscopically determined radial velocities. The phase difference of the stellar eclipses
constrains well the quantity $ e \cos\omega / \sqrt{1-e^2}$; so the eccentricity $e$ and the
longitude of periastron, $\omega$, are constrained by the Kepler photometry. Also, the ratio of the
difference in durations (secondary eclipse minus primary eclipse) to the sum of the two durations
equals $e \sin\omega$, to a good approximation that the orbital inclination is $\sim 90\arcdeg$. 
With Doppler boosting, the Kepler photometry provides the equivalent of a single-lined spectroscopic
binary: the radial velocity of the brighter star as a function of orbital phase. The latter also
constrains $e$ and $\omega$ and the phase of periastron passage, along with a measure of the radial
velocity semiamplitude.  For stars of similar brightness, the Doppler boosting effects of the two
stars will tend to cancel. Of course, the traditional light-curve analysis of the eclipses provides
an estimate of the relative brightnesses of the two stars in an EB. For those eclipsing
binaries with one star much brighter than the other, and if that brighter star is photometrically stable or
at least predictable as in the case of \object{KIC 4862625}'s rotational modulation, Doppler boosting curves
from Kepler photometry may provide radial velocities adequate for estimating the mass function of
the system, and other parameters of the EB.

In this work, we have demonstrated that the stellar radial velocities can be measured
either with a spectrometer or a photometer. We give priority to the spectroscopic technique
because of its well-calibrated heritage.
For comparison we measured
$k_1 = 16.7\pm 0.5$ and $\omega = 222\arcdeg \pm 2\arcdeg$ from Doppler boosting
in the Kepler light curve. We estimated the maximum-likelihood values of $k_1$ and $\omega$
from all of the photometry. However, because correlations are apparent in the residuals
of the Doppler boosting curve, induced by the 2.63-day averaging window during detrending, we estimated the
uncertainties of $k_1$ and $\omega$ from subsets of points selected to
be independent from each other, in steps of phase equal to 2.63 d / 20 d.
The photometrically-determined value of $\omega$ is consistent with
that determined from shifts of spectral lines, $\omega = 220.2\arcdeg
\pm 3\arcdeg$. Formally, the photometrically-determined value for
$\omega$ is more precisely determined than the spectroscopically-determined
value.
The value of $k_1$ is prone to systematic
error; it is directly proportional to the boost factor $B$ estimated
from the overlap integral of the stellar spectrum and the Kepler
bandpass.  The systemic velocity $\gamma$ is indeterminate from
Doppler boosting.

\subsection{Stellar Rotation and Star Spots}

From the broadening of the spectral lines and the period of amplitude modulations
in the light curve, we infer rotation of star A and determine its
radius,
\begin{equation}
R_A = { { p_{rot} \Vsini} \over { 2 \pi \phi} },
\end{equation}
where $\phi$ is a factor of order unity that would account for differential
rotation and any systematic errors in \Vsini, such as $\sin i_* < 1$. 
With $\Vsini = 31\pm2\,$km\,s$^{-1}$, ${\rm p_{rot} = 2.6382\pm 0.0037}$ d,
and $\phi = 1$ the stellar radius $R_A = 1.62 \pm 0.1 R_\odot$, in agreement
with the estimates from the spectral analysis (Section \ref{sec:spectralanalysis}) and
the photodynamical model (Section \ref{sec:results}). With $R_A$ determined from
rotation, the density of star A implied from the light curve and the mass function
from the radial velocity semiamplitude, we derive a mass $M_A = 1.2 \pm 0.2 M_\odot$,
again in agreement with estimates from the other two methods. We also estimated $M_A$ using the 
predicted velocities of the primary and the planet during the nine transits from the best-fit simulations.
Each of the nine events is scaled for the respective velocities, such that they have 
approximately the same width. All nine are then fit together. The derived parameters of the primary star 
agree with the above mentioned values within their uncertainties.
Because $M \propto R^3$, the fractional uncertainty in stellar mass is three times larger than the
stellar radius' fractional uncertainty, which is dominated by the uncertainty
in \Vsini\ and any bias implicit in $\phi$.

A curious situation is possible with \object{KIC 4862625}. 
Star A's peak orbital velocity transverse to the line of sight is $\sim$19 km s$^{-1}$
and its rotational velocity is 31 km s$^{-1}$ (Section \ref{sec:spectralanalysis}).
Assuming that star A's rotation is prograde with respect to the orbital motions
of star B and the planet, and the axis is in the plane of
the sky, i.e. sin i = 1, and that the transit is a central one, 
the transverse velocity of a star spot (or plage) could briefly match
that of the planet ($\sim$46 km s$^{-1}$)! 
Planetary transits could exhibit asymmetric shapes due to the nearly matched
transverse speeds. \object{KIC 4862625}'s transit events 1, 3, and 5 in particular appear asymmetric, with
egress being more abrupt than ingress.  We are unsure whether the asymmetry is astrophysical or an artifact of
the detrending.
Because they are grazing, the primary eclipses of \object{KIC 4862625} may not
exhibit similar asymmetries if spots (or plage) are not prevalent near
the pole of star A, and in any case the projected rotational speeds
will be small at the star's poles.
Silva (2003, 2008) and Nutzman, Fabrycky, \& Fortney (2011) have analyzed
star-spot induced asymmetries in transit light curves but did not explicitly consider
the possibility of a star spot over-taking a planet. More typically, e.g. for a 3-day
period planet transiting a solar-type star rotating every 30 days, the 
transverse velocity ratio (planet/star) is $\sim$100.
Gravity darkening is another mechanism for which very rapid stellar rotation can induce subtle asymmetries
in transit light curves (e.g. Barnes, Linscott, and Shporer 2011).
\section{Diagnosing a System}
\label{sec:puzzle}

Initial diagnosis of a single transiting planet that orbits a stellar binary can be challenging,
particularly if orbital period is long and the planet transits only one star.
There are many system parameters and potentially only a few observational constraints.
In some cases the existing data will not fully constrain the system.
Given this potential complexity, it is useful to understand the sequence of analysis steps
that build understanding of a newly discovered system.
With this in mind, we describe the clues we used to diagnose the \object{KIC 4862625} system.

First, we worried that \object{KIC 4862625} could be an astrophysical false positive. The aperiodicity
of the nine planetary transits disproves the hypothesis of a background eclipsing binary mimicking
a transiting CB planet. Superficially, the false-positive of a dilute EB mimicking planetary transits of a single
star (Brown 2003) has an analogy in CB planets, namely that a dilute eclipsing triple star in proximity to
an EB star could in principle mimic some of the characteristics of a transiting CB planet.
First, however, the chance proximity on the sky of a
double star and a triple star will be much rarer than that of a single star and a double star (Lissauer et al. 2012).
Second, while one could contrive a dilute triple star to mimic the aperiodic centers of a few transit-like events, the
durations in general would not match also, because for a CB planet, the transit durations depend
critically on the characteristics of the EB (Eq. \ref{eq:durations}). We conclude that
a third body orbits the EB star \object{KIC 4862625}.

We next analyzed the stellar binary, which has best observational constraints.
The Kepler light curve folded on the 20 day period of the stellar binary (Figure~\ref{fig:lc})
has a primary eclipse that is 1.3\% deep and a secondary eclipse that is 0.1\% deep.
Primary eclipses are more V-shaped than U-shaped,
suggesting that a smaller secondary grazes a larger primary.
Secondary eclipses have a flat bottom with gradual ingress and egress,
suggesting that the limb of the primary fully occults the secondary.
The phase difference between primary and secondary eclipses indicates an orbit with significant eccentricity.

Next we examine high-resolution spectra of the system.
Spectral features are roughly similar to the Sun, except that the lines are shallow and broad.
Cross-correlation with a template yields radial velocity shifts consistent with the light curve period of 20 days.
Line widths imply $\Vsini=31$ \kps, which is typical for spectral types slightly earlier than the Sun.Ê
As expected, lines of the low-mass secondary are not detected in the optical spectra.

Next we examine the nine planetary transits that exist in publicly available Kepler data.
First we checked the data quality flags to verify that all data are valid during the transits.Ê
Using the PyKE software\footnote{http://keplergo.arc.nasa.gov/PyKE.shtml} with custom apertures to analyze Kepler target Pixel files,
we confirmed that the nine transit events come from the central target pixels.
We measured no significant centroid shifts during the transits,
which rules out certain astrophysical blending scenarios.
We measured transit start and end times by fitting U-shaped functions.
We then calculated transit midpoint times and durations.

The cadence and duration of the transits suggest that all involve the primary star.
The cadence of the planetary transits yields an apparent orbital period of approximately 136 days.
The time differences between successive transits is 136.6, 136.7, 135.9, 133.2, 135.9, 136.7, 136.4 and 134.7 days. Such regularity is not anticipated for a circumbinary system,
where a single object transits or eclipses a ``moving target'' (Orosz et al. 2012).
The regular cadence of transits is not caused by simple commensurability,
given the 20 and 136 day periods in the system.

We use the durations of the transits to constrain the parameters of the stellar binary. 
Transit duration depends on chord length and on transverse velocity of the stellar primary relative to the circumbinary object (Schneider \& Chevreton (1990). During transits, transverse velocity of the planet is always fairly similar, but transverse velocity of the occulted primary (dependent of the mass ratio of the two stars) changes direction and amplitude throughout its orbit. The amplitude of the transverse velocity of the primary decreases to a minimum at orbital turning points. For eccentric stellar binaries, amplitude also decreases with increasing binary separation. Finally, transit durations are longer when the primary is in the portion of its orbit where transverse velocity of the primary and planet are aligned, and shorter when the two velocities have opposite signs.

With these factors in mind, we now interpret the nine observed transits, ordered by increasing duration.
Transit 2 has the shortest duration because the transverse velocities are oppositely directed 
and the primary is far from a turning point.
Transit 1 has a slightly longer duration because the primary is slowing as it approaches a turning point.
Transit 3 is near the other turning point, where binary separation is larger, further reducing relative velocity.
Transit 4 has a relatively long duration because the transverse velocities are now aligned,
though the primary is near the turning point where binary separation is large.
Transit 5 has very long duration because velocities are still aligned
and the primary is near the turning point where binary separation is small.
Transit 9 has a very long duration because the star and the planet travel ``side by side''.
Because transit durations are shortest near primary eclipse,
the planetary orbit must be prograde relative to the stellar binary.

\begin{figure}
\centering
\plotone{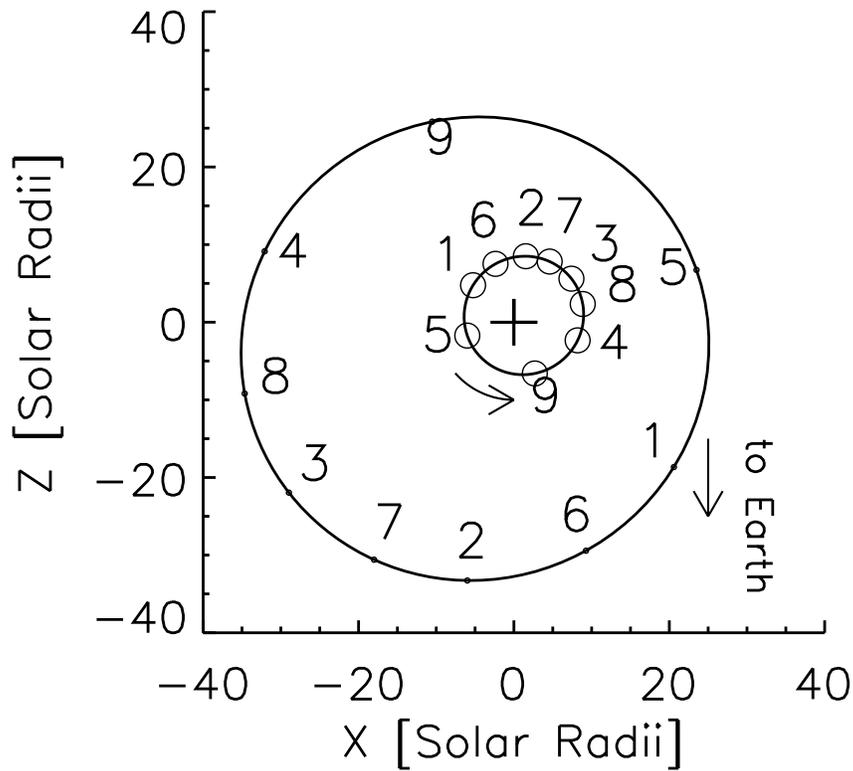}
\caption{
Scaled view of the binary system of {\protect \object{KIC 4862625}} viewed perpendicular the orbital plane. The inner ellipse shows motion of the primary, while the outer ellipse shows the secondary. The two stars orbit the barycenter of the system (cross symbol) in a counter-clockwise direction. Numbers along the orbits indicate the configuration of the binary system at the times of the planetary transits. At these times, the planet and the primary star align along the y-axis. The planetary orbit is outside the scale of the figure. The diagram is to scale.
\label{fig:top_view}}
\end{figure}

\begin{figure}
\centering
\plotone{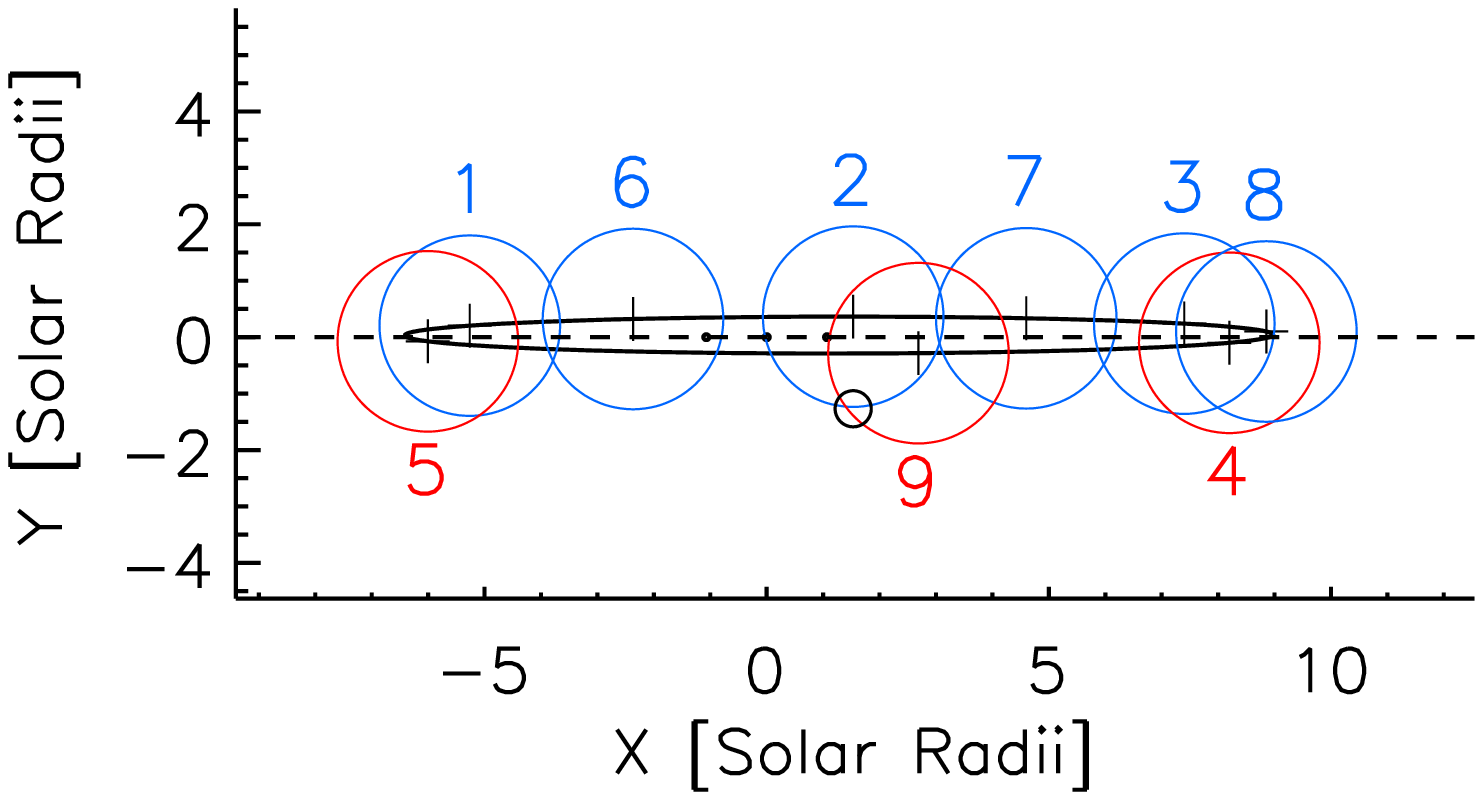}
\caption{
Schematic scale diagram of {\protect \object{KIC 4862625}}. 
The diagram is to scale, with solar radii indicated on the axes.
The positions of star A in its orbit (large circles) are plotted 
at the times of the nine planetary transits (labeled, large circles centered on black $+$ symbols). Red color indicates the position of the primary star on the near side of its orbit and blue color -- on the far side. Star B is shown to scale (small circle, arbitrarily positioned below star A at event 2). The planet also is shown to scale (smallest circles);  the outer two corresponding to positions $-$0.25-d and $+$0.25-d with respect to the center position, which, has been arbitrarily centered at the origin.
\label{fig:side_view}}
\end{figure}

To quantify constraints provided by observed transit times and durations,
we calculated the locations of all bodies in the system as a function of time.
Figure \ref{fig:top_view} and \ref{fig:side_view} are schematic scale drawings of the EB system. 
Figure \ref{fig:side_view} illustrates the sizes of the three objects (star A, star B, and the planet) relative to each other and to the barycentric orbit of star A, which stretches only a few stellar radii across. 
The schematics also illustrate the positions of the stars at the epochs of each of the nine planetary transit events. Because we have approximated the transit chord as the diameter of star A for all nine planetary transit events, we have not illustrated the orbit of the planet in the schematic diagrams.
The primary has a very similar projected position during transits 1 and 5 and, half a binary period later, during transits 4 and 8.
Because of this near coincidence, the time difference between transit 1 and 5
is almost exactly four times the 135.59 day orbital period of the planet. 

Next, we inspect event 4, also a relatively long duration event. At the phase of event 4, the primary
star is near the opposite turning point of its orbit from events 1 and 5.
Note event 2 is near primary eclipse, when the two star's relative motions are relatively fast, and
the primary star is traveling right to left, i.e. in the opposite direction to the planet, 
if in our scenario the planet is traveling in a prograde orbit (left to right in this model).
Again, event 2 is deeper than secondary eclipse, and event 2's shape is not a square wave, so we
consider event 2 is again a transit of the planet across the primary star. 

Transverse velocity of the primary is a function of true anomaly, argument of periastron, and eccentricity.
Using the nomenclature of Hilditch (2001), transverse velocity $V_x$ is given by

\begin{equation}
V_{x} =  - (\frac{M_2}{M_{bin}}) (\frac{2 \pi G M_{bin}}{P_{bin}})^{1/3} \frac{(e\sin\omega + \sin(\theta + \omega))}{(1-e^{2})^{1/2}},
\label{eq:vx}
\end{equation}
where $M_{bin}=\Mpri + \Msec$, $\theta$ is true anomaly, $\omega$ is the argument of pericenter and {e} is the eccentricity of the binary star. 

Using the formalism of Schneider \& Chevreton (1990), and assuming a circular orbit for the planet, we obtain the following expression for transit duration $t_{\rm dur}$:

\begin{equation}
t_{dur} =  \frac{AB}{1 + ACx}
\label{eq:durations}
\end{equation}
where
\begin{equation}
\begin{split}
A = (M_{bin})^{-1/3} \\
B = 2R_c (\frac{P_p}{2 \pi G })^{1/3} \\
C = - f(m) (\frac{P_p}{P_{bin}})^{1/3} (1-e^{2})^{-1/2} \\
x = (e\sin\omega + \sin(\theta + \omega))
\end{split}
\label{eq:durations2}
\end{equation}

In Eq. \ref{eq:durations2}, we use the mass function $f(m)$ obtained from radial velocities
to substitute secondary mass ($M_{2}$) for binary mass ($M_{\rm bin}$); $x$ is calculated from 
the known $\theta$ of the binary star at the time of each planetary transit and $t_{dur}$ is measured
for each of the nine events. 
The mean period of the planet ($P_{p}$) is known from the cadence of transit times.
Assuming transit chord ($R_{c} = \Rpri + R_{planet}$) is the same for every transit, we fit for the coefficients $A$ and $B$ ($C$ is known to the precision of $P_{p}$). 
The inferred value of parameter $B$ in Eq. \ref{eq:durations} constrains stellar radius.
Long duration transits will deviate slightly from Eq. \ref{eq:durations},
if velocity of the stellar primary changes significantly during the transit.
We caution that very small changes in observed transit durations
can have significant effects on parameters derived from $A$ and $B$.

\begin{figure}
\centering
\plotone{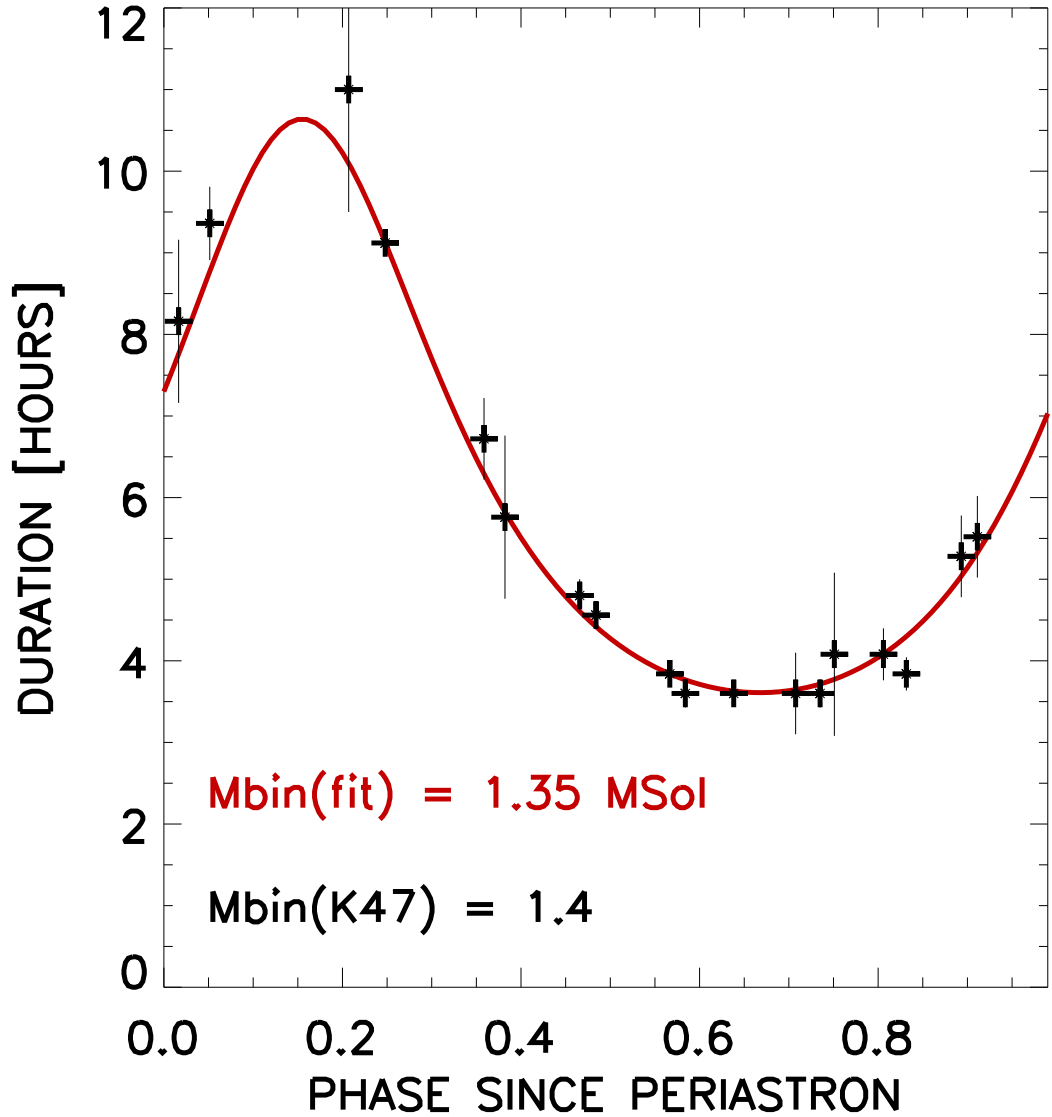}
\caption{
Duration versus phase for the Kepler-47b transits. The red curve represents a fit from Eq. ~\ref{eq:durations}. The analytic expression recovers the parameters of the binary star very well -- $M_{bin, calc} = 1.35 \Msun$ vs $1.4 \Msun$ and $R_{prim} = 0.87 \Rsun$ vs $0.96 \Rsun$ respectively, provided by Orosz et al. 2012.
\label{fig:dur_K47b}}
\end{figure}

\begin{figure}
\centering
\plotone{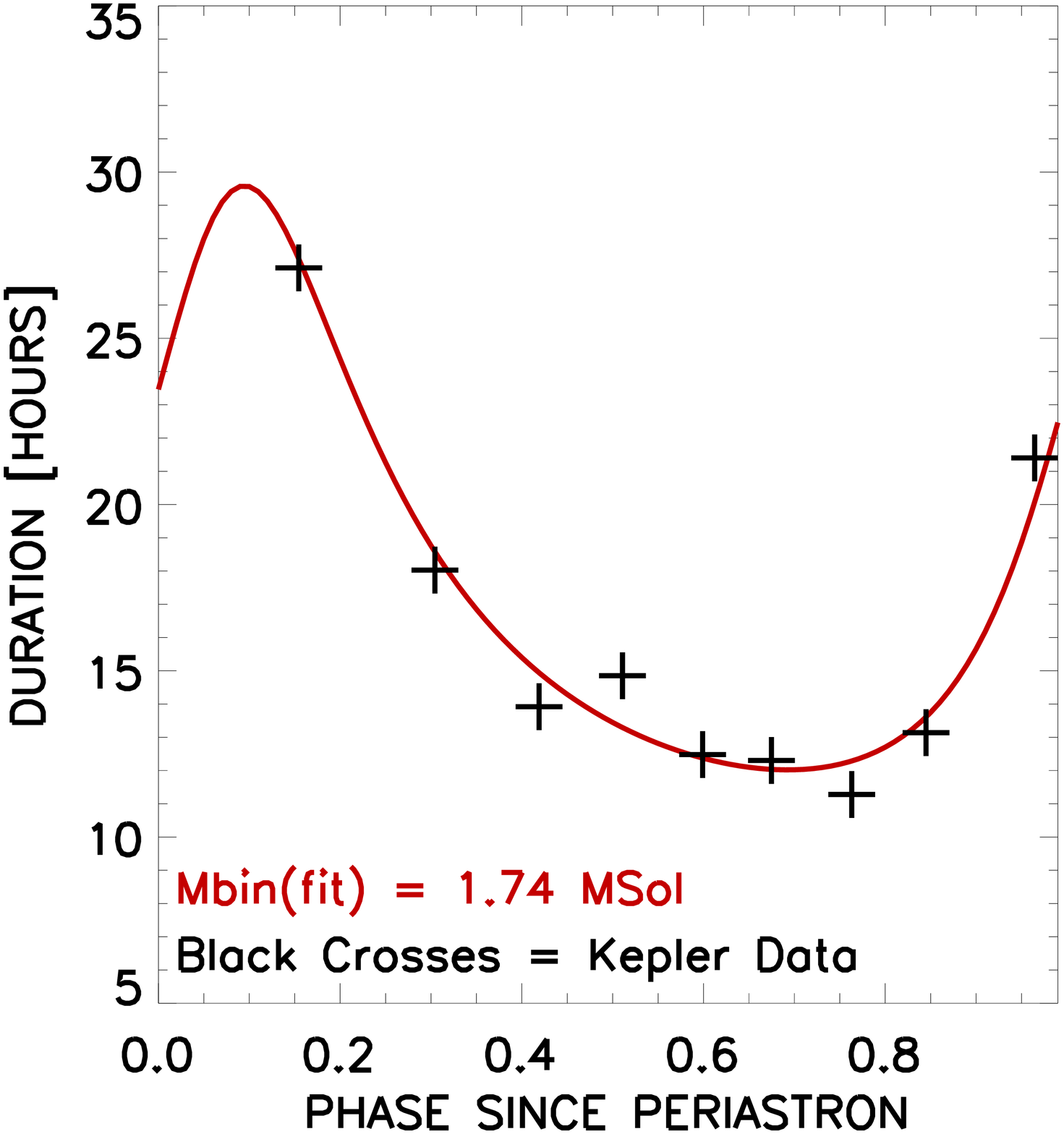}
\caption{
Similar to the previous figure but for {\protect \object{KIC 4862625b}}. The uncertainties in the measured durations are smaller than the size of the symbols.
\label{fig:dur_486}}
\end{figure}

To validate Eq. \ref{eq:durations}, we fitted observed transit durations for Kepler-47b,
using transit parameters in Orosz et al. (2012). The fit shown in Figure \ref{fig:dur_K47b} yields $M_{bin}=1.35$ \Msun and $\Rpri=0.87$ \Rsun,
approximately equal to the $1.4\pm0.05$ \Msun\ and $0.96\pm0.017$ \Rsun\ deduced by Orosz et al. (2012). 
Figure \ref{fig:dur_486} shows an analogous fit for \object{KIC 4862625b}.
We obtain an $M_{bin}=1.74$ \Msun\ and $\Rpri=1.7$ \Rsun.
The analytic curve indicates that future transits may be as long as 30 hours!
To assess uncertainties, we created simulated observations
for a mock system like \object{KIC 4862625} (Model 1 in Section \ref{sec:results}).
We integrated the mock system for 9 planetary orbits, using a time-step of 14.4 min, to 
calculate durations and times of the planetary transits and then fit for A and B in Eq. \ref{eq:durations}.
As with Kepler-47b, the inferred binary mass of 1.76 \Msun\ is $\sim 5$\% smaller than the
expected value of 1.85 \Msun.
The chord length calculated from $A$ is 1.68 \Rsun, compared to the input value of 1.7 \Rsun.
The fits to the planetary transit durations using the analytic functions provide a good starting point
for the more refined fits presented in Section \ref{sec:results}.

In the existing data, depths of the nine observed transits appear to change with time.
Perhaps this is simply an artifact of our detrending procedure.
Alternatively, star spots that perturb the light curve may also affect apparent transit depth.
Finally, changes in transit depth could be a manifestation of a planetary orbit with an inclination different than 90 degrees.

Numerically integrating an inclined planetary orbit forward in time,
the transits fade away, cease for a period, and then return after a large number of orbits.
The larger the inclination, the faster the evolution.
Given that we have observed 9 transits so far,
we assume that the chord lengths, transit depths, and planetary inclination are constant.

For completeness, we note that the observed dependence of the durations of the nine transits on the phase of the binary rule out a retrograde orbit for the planet.
Such an orbit should exhibit an opposite trend in a duration versus phase diagram, namely short durations near secondary eclipse and longer ones close to primary eclipses. 

\section{Planetary Transits}
\label{sec:results}

\subsection{Methodology}

Combined light curve and radial velocity measurements of a circumbinary planet can be characterized by 17 parameters: five orbital elements of the binary ($a_{bin}$, $e_{bin}$, $\omega$, $\Omega$, $i$), six osculating orbital elements of the planet ($a_3$, $e_3$,  $i_3$, $\omega_3$, $\Omega_3$, $\phi_0$), three masses (\Mpri, \Msec, $m_3$), and three radii (\Rpri, \Rsec, $r_3$). Exhaustively searching a space with 17 dimensions and nonlinear parameters is impractical, so we make simplifying assumptions that allow us to obtain a reasonable solution. As discussed in Section RV, radial velocity measurements yield $e$, $\omega$, and the stellar mass ratio. The precise Kepler light curve yields the ratio of stellar radii, which depends on derived impact parameters for eclipses that may be grazing. We assume the nine observed transits all occult the primary star, so radius of the secondary star drops out of the system. For simplicity, we assume that the planet has negligible mass and orbits in the plane of the sky ($i_3=90$ deg). The former assumption is based on the measured ETVs, discussed in Section ETV and expanded below. The latter relies on the hypothesis that the inclination of the planet should be certainly larger than that of the binary for otherwise it would not be seen in transit. 

With these approximations, the four remaining free parameters are \Mpri, $a_3$, $e_3$, and $\omega_3$. Finding optimal values for these four parameters still requires a very fine numerical grid. For example, a 1\% change in mass of the primary star, or in the semi-major axis of the planet can dramatically change the dynamical evolution of the planet, affecting arrival times of observed transits by hours or even days. Arrival times would also be affected if the planet is massive, rather than a massless test particle. 

\begin{figure}
\centering
\epsscale{0.8}
\plotone{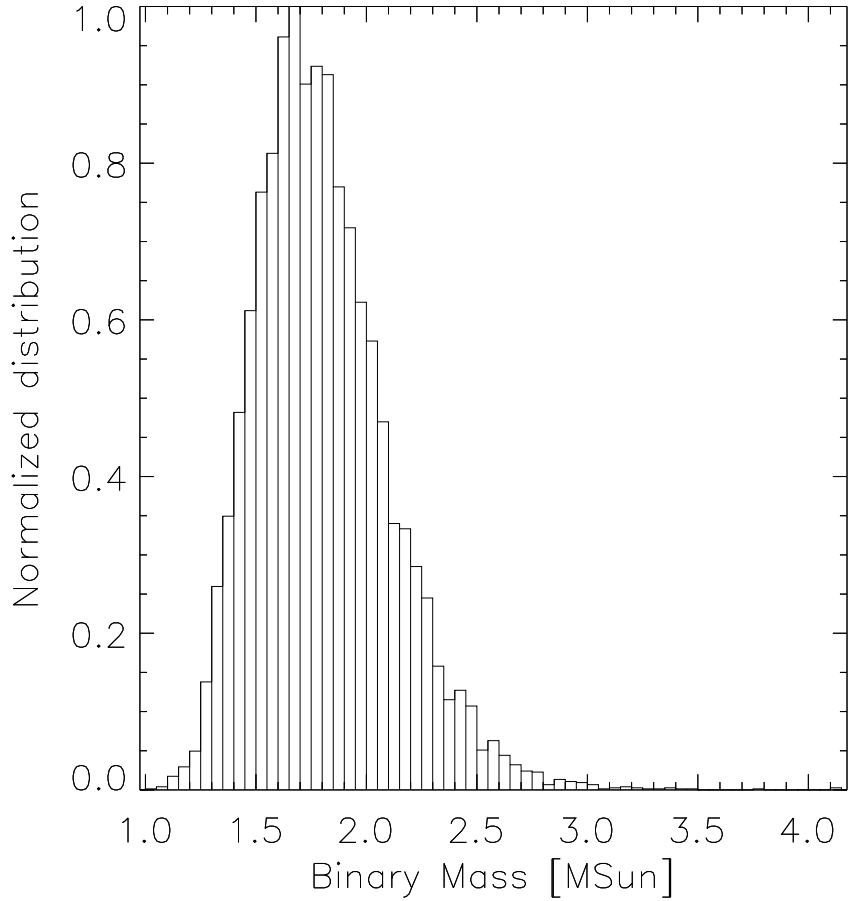}
\caption{
Mass of the binary ($M_A + M_B$) derived using the results from the RV measurements and applying Eq. \ref{eq:durations} to ten thousand iterations of the nine planetary transit durations, scrambled and with added normalized noise with $\sigma$ = 1 hour. We used this distribution to set the limits of our search space for the primary mass in our dynamical model of {\protect \object{KIC 4862625}}.
\label{fig:scrambled}}
\end{figure}

Transit durations (shown in Table \ref{tab:future}) have formal uncertainties that do not necessarily account for astrophysical variations in the light curve. To assess how sensitive binary mass is to individual transit durations, we generated and analyzed a set of perturbed observations. We scrambled the nine observed transit durations and added a normally distributed perturbation with a standard deviation of 1 hour (2 long-cadence data points). We then used Eq. \ref{eq:durations} to evaluate the binary mass, obtaining the results shown in Figure \ref{fig:scrambled}, where we plot the normalized distribution of best-fit models as a function of the mass of the binary. The binary mass distribution has 68\% of the values in the range 1.4 to 2 \Msun. In subsequent analysis, we constrain the binary mass to be in this range.

A simple Keplerian solution cannot reproduce observed planetary transit times because the central potential varies as stars in the binary orbit each other. Instead we use a three-body numerical integrator with hierarchical Jacobian coordinates, as will be further described in Section \ref{sec:dynamics}. We specify initial conditions (and reference epoch) with respect to the midpoint of the first transit, an arbitrary yet convenient definition. The coordinates of the three bodies are defined with respect to the barycenter of system, which is the barycenter of the stellar binary when planet mass is negligible. We use time-dependent, osculating Keplerian elements to calculate $a_3$, $e_3$, $i_3$, and $\omega_3$ (Doyle et al. (2011)). We use the numerical integrator to compute spatial coordinates for the two stars and the planet. Transits intervals occur when the projected distance between the primary star and planet centers is less than the sum of their radii. Note that parameters cited in this section are instantaneous values for the reference epoch.

First, we used a coarse time step (one third the shortest transit time) to calculate dynamical solutions and corresponding transit midpoint times for a grid of parameter values.
The grid step sizes were 0.01 \Msun\ for \Mpri, 1 \Rsun\ for $a_3$, and 5 deg for $\omega_3$.
Changes of 1 \Rsun\ in $a_3$ cause significant changes in transit midpoint times.
Parameter ranges in the coarse grid were 1.4 to 2.0 \Msun\ for \Mpri, 0 to 0.2 for $e_3$, and 0 to 360 deg for $\omega_3$.
Next, we used a fine time step (0.01 days) to calculate dynamical solutions for a much finer grid of parameters around the best solutions in the coarse grid.
The fine grid includes $r_1$ as a fifth parameter and calculates transit durations in addition to transit midpoint times.
Comparing computed and observed midpoint times and durations for the nine transits yields a set of plausible models.

\subsection{Results}

\begin{figure}
\plotone{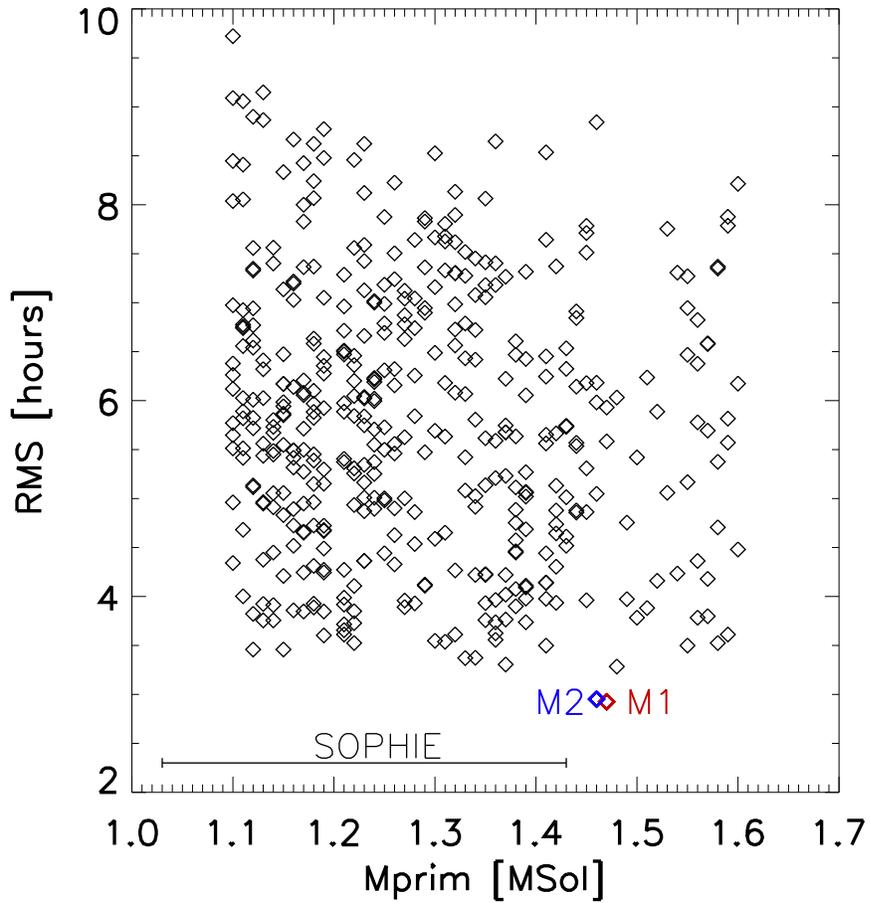}
\caption{
Best-fit dynamical models of {\protect \object{KIC 4862625}} predicting the midpoint times of transits 2 through 9 within an RMS of $\sim7$ hours. Models with different parameters produce similar solutions, all predicting planetary transits that deviate by only a few data points from the observed values, a small error margin on the orbital scale of the planet. The decrease in the number of ``hits'' as the mass of the primary increases from 1.1 to 1.6 \Msun is not a systematic effect but is in fact real -- there are many more good ``hits'' for smaller $M_{prim}$.
\label{fig:dynhits}}
\end{figure}

Even searching in a restricted parameter space, we found multiple solutions that are consistent with the nine planetary transit observations.
We adopted a goodness of fit metric that is the root mean square (RMS) of the sum of observed minus calculated (O$-$C) midpoint times for transits 2 through 9.
The midpoint time for transit 1 always has zero residual because it defines the zero point of time in our calculations.
Figure \ref{fig:dynhits} shows goodness of fit for every dynamical model in the grid that matches observed transit midpoint times to better than a RMS of $\sim10$ hours.
Each point in the figure represents a unique combination of \Mpri, $a_3$, $e_3$, and $\omega_3$.
Despite a uniformly sampled grid, the number of  solutions better than RMS of $\sim10$ hours decreases with increasing \Mpri, suggesting that lower \Mpri\ is more likely. We note, however, that even though almost half of all solutions fall within \Mpri between 1.1 \Msun and 1.3 \Msun, most of them have high RMS and are not very likely.

Within the limitations of our grid space search, two solutions stand out from the rest as having an RMS O$-$C to within 3 hours. The two models, labeled M1 and M2 in Figure \ref{fig:dynhits}, reproduce the observed planetary transits equivalently well. Further supporting the existence of {\emph a} solution near these two models stems from them having very similar parameters -- both indicate a primary star of $\Mpri=\vMstarA\Msun$ and a planetary orbit with a semi-major axis  $a_3=0.64$ AU, and $e_3\sim0.1$. Combined with observational constraints described in previous sections, this implies $\Msec = \vMstarB\Msun$, $\Rpri = \vRstarA\Rsun$ (consistent with our estimates from the spectra and the stellar rotation), $\Rsec = \vRstarB\Rsun$, and $P_3=138.5$ days. After global minimization, individual transits are inspected for significant inconsistencies between observed and model light curves. Agreement is very good for the two models, but there are subtle differences -- the model transits oscillate around the observed ones to within a few data points. Further examination shows that M1 clearly stands out as M2 consistently under predicts the mid transit times of events 5 through 9 by up to an hour.

\begin{figure}
\centering
\epsscale{1.0}
\plotone{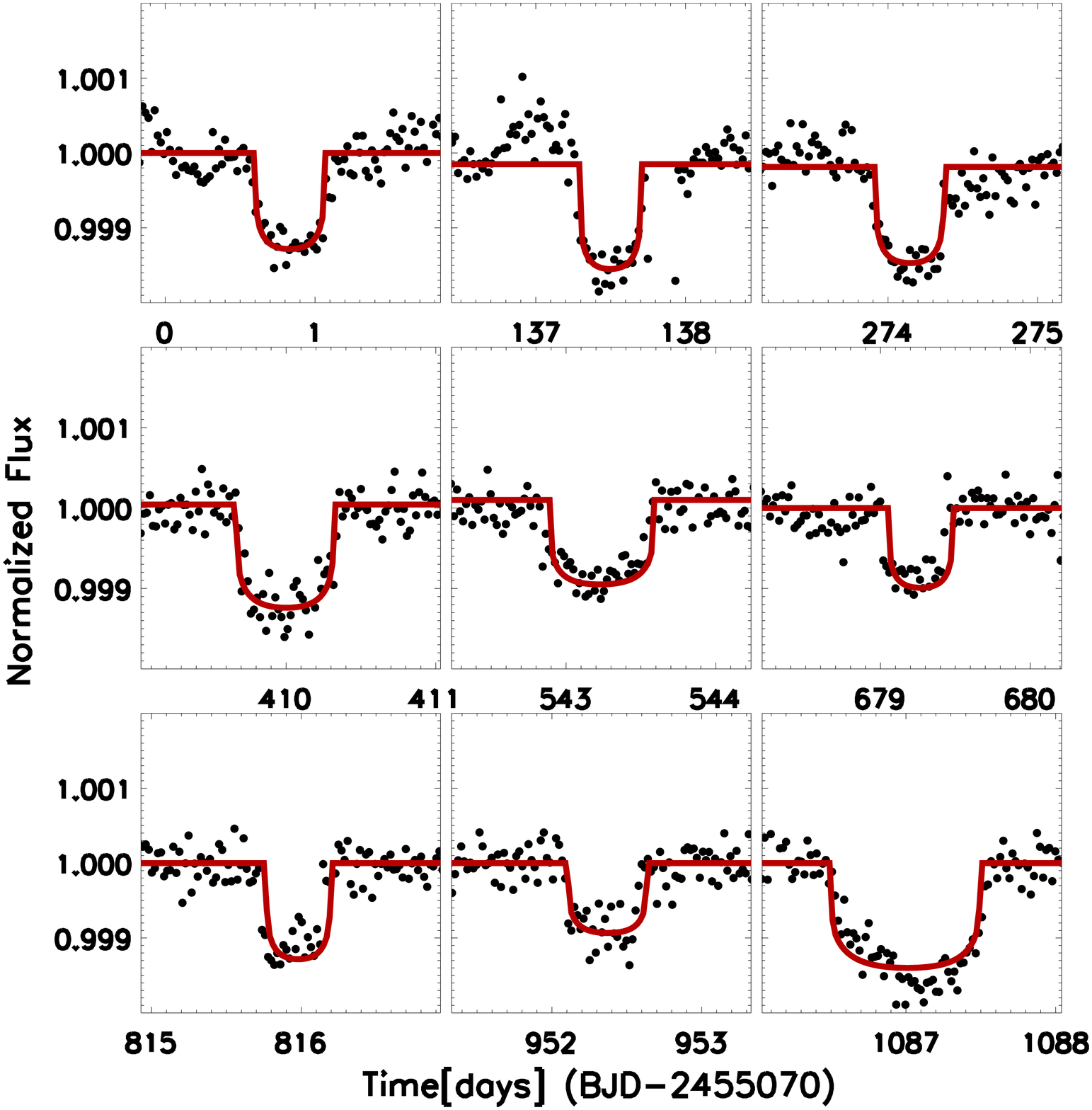}
\caption{The best-fit photodynamical model of {\protect \object{KIC 4862625}} predicting the center times for transits 2 through 9 to within RMS of three hours. The model supports a primary star of $\vMstarA\Msun$ and $\vRstarA\Rsun$ radius. All simulated planetary transits fit to within 2 data points of the measured values of both the durations and the mid transit times. Future events will allow us to further constrain the characteristics of the system. The orbital and physical parameters of the system are outlined in Table \ref{tab:planmods}.
\label{fig:photodynamics}}
\end{figure}

Table \ref{tab:planmods} lists the best-fit parameters for model M1; the photodynamical solutions to the planetary transits are shown on Fig.~\ref{fig:photodynamics}. The model reproduces well the observed phase-dependance of the durations, with event 2 predicted to be shortest and event 9 -- longest. We caution that the nine observed transits are not sufficient to completely rule out any plausible solutions. Other equivalent or even better solutions may exist, given the coarseness of our initial grid search. We were unable to match observations with a circular orbit ($e_3=0$), which is not surprising given that orbital elements evolve continuously. An orbit that is initially circular will change with time, especially when the planet is relatively close to the stellar binary (e.g., \object{KIC 4862625}).

To estimate the uncertainty of the derived \Mpri, we examined the distribution of solutions as a function of $\Mpri$ that have an RMS of less than 6 hours ($\sim2\sigma$ above the best-fit value). Nearly half of the solutions shown on Figure \ref{fig:dynhits} comply with this merit criteria, of which 92 fall within $\Mpri$ between $1.3\Msun$ and $1.5\Msun$. The distribution is double-peaked with a taller one centered on $\Mpri\sim1.15\Msun$ and another, smaller peak near $\Mpri\sim1.4\Msun$. The low-mass solutions, however, while producing a good overall RMS, consistently result in larger O$-$C deviations for individual events compared to the higher-mass solutions. Thus, we focus on the $\Mpri\sim1.4\Msun$ peak, where half of the solutions fall within a range of $\Mpri$ between $1.32\Msun$ and $1.48\Msun$ with an average of $1.4\Msun$. Similarly, we constrain the semi-major axis of the planet and the argument of periastron but not the eccentricity of the orbit where the best-fit values were uniformly spread. We report the values in Table \ref{tab:planmods}. At the time of submission, we predicted durations of planetary transits that would follow the first five. Transit 9 we predicted to have a very long duration, and indeed it does (nearly 30 hours), allowing us to significantly improve the constraints on the stellar masses. Transits 6, 7, and 8 we predicted to have short durations similar to many other transits, providing only limited additional information about the mass of the binary star. 

\begin{figure}
\centering
\epsscale{0.75}
\plotone{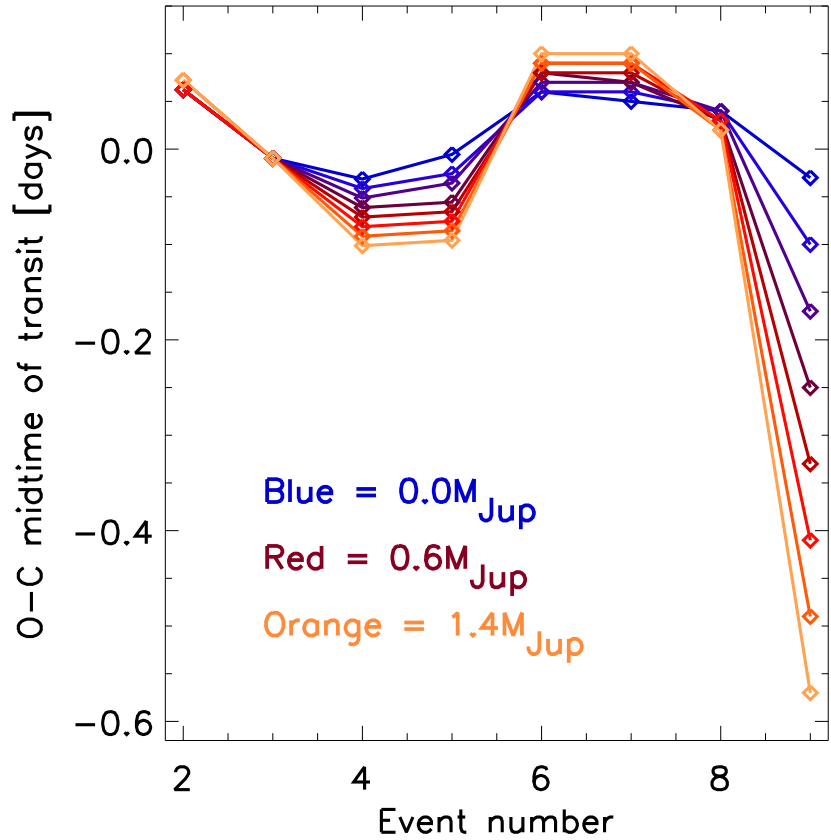}
\caption{
The effect of a non-zero planetary mass on the predicted times of transit of {\protect \object{KIC 4862625}}. Keeping the best-fit parameters of the circumbinary system constant, an increasingly more massive planet introduces significant deviations relative to the best solutions for a massless planet. While it is unlikely that the mass of the planet is larger than a few \Mjup, it adds a significant complication to the best-fit models, again indicating the non-unique nature of the solutions.\label{fig:planmass}}
\end{figure}

We also investigated how non-zero planet mass (1 to 10 \Mjup) affects transit midpoint time and the precise cadence of stellar eclipses by perturbing Model 1. Using the formalism of Borkovits et al. (2012), we estimate that light travel time effects due to a 10 \Mjup\ planet would produce timing changes with an amplitude of only a few seconds. On the other hand, dominant three-body dynamical interactions would produce an O$-$C variations with an amplitude of $\sim3.5$ minutes, which is approximately twice the measured value (Figure \ref{fig:etv}). Thus, the mass of a coplanar planet should not exceed $\sim5$ \Mjup. Figure \ref{fig:planmass} shows how predicted transit midpoint times change as planet mass increases from 0 to 1.4 \Mjup. A more massive planet tends to arrive earlier for events 4, 5 and 9 and late for 6 and 7, where the deviations are most pronounced for event 9. We tested several other models and found they all show similar behavior, indicating that while Saturn-mass planets and smaller can be safely approximated as test particles for dynamical purposes, a more massive planet would need to be properly accounted for in numerical integrations of \object{KIC 4862625}.

\subsection{Search for additional transits}

We visually examined the light curve of \object{KIC 4862625} for evidence of additional transits.
Despite significant variability, there are a few features that are reminiscent of shallow planetary transits.
For these features, data flags are normal and centroids have no significant shifts.
Some of these extra features occur close in time to the nine main transits, most notably around day 5615 (BJD-2450000) when the dynamical model predicts the planet is near the secondary star, but also near days 5045 and 5069, where the two are distinctly apart. The small size and faintness of the secondary, however, means we do not expect to see a quaternary transit (see Section \ref{sec:eblc}). Others (near days 5258, 5269, 5298, 5312, 5415; BJD-2450000) cannot be easily associated with the circumbinary planet \object{KIC 4862625b}. 
\section{Dynamical analysis and orbit stability}
\label{sec:dynamics}

We have carried out a dynamical analysis of the KIC4862625 system within the framework of the three-body problem. Such a dynamical system is well known to possibly exhibit complex dynamical behavior. In particular we have carried out a dynamical analysis of the planet around the binary pair in order to detect chaotic regions, often associated with mean-motion resonances (MMRs), in the orbital parameter space of the planet. 

We have applied the MEGNO\footnote{Mean Exponential Growth of Nearby Orbits} factor \citep{cincottasimo1999, cincottasimo2000, cincottaetal2003}. This numerical technique is efficient in distinguishing between chaotic and quasi-periodic and has found widespread application within dynamical astronomy and the dynamics of multi-body extrasolar planets \citep{GozdziewskiMEGNO2001,Gozdziewski2008,Hinse2010}.

Chaotic orbits are usually (but not always) attributed to unstable orbits. For a quasi-periodic time evolution of the system the dynamics is regular and characterized by only a few fundamental frequencies often associated with stable orbits. However, in order to be precise, quasi-periodic/stable orbits can only be quoted as stable up to the considered integration time. Knowledge on the subsequent dynamical evolution of the system is still hidden to the experimenter. In this work we have experimented with various integration length and found an integration time scale that is long enough to detect the most important mean-motion resonances close to the osculating orbital elements of the transiting planet KIC4862625b.

\begin{figure}
\centering
\epsscale{0.8}
\plotone{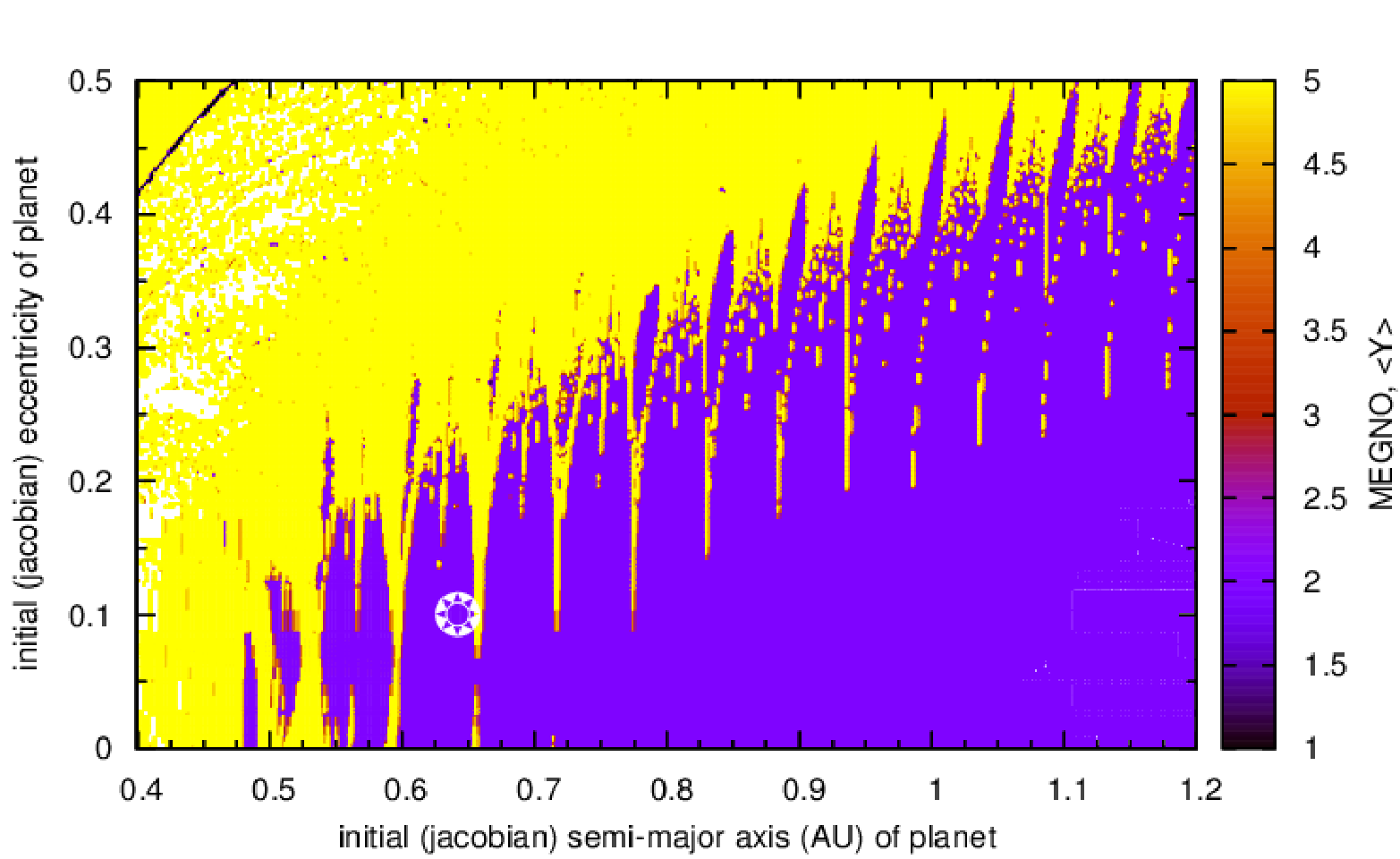}
\plotone{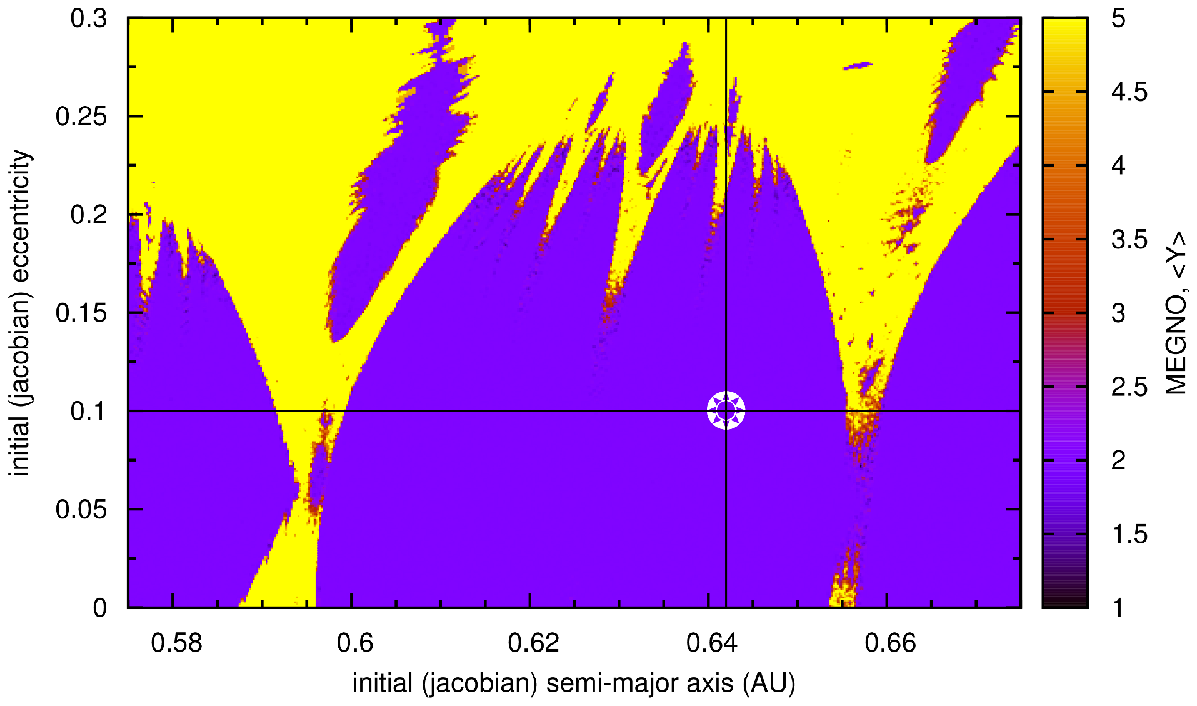}
\caption{Upper panel: MEGNO map of {\protect \object{KIC 4862625b}}. The mass of primary and secondary component were set to $1.47~M_{\odot}$ and $0.37~M_{\odot}$, respectively. The binary semi-major axis and eccentricity was set to 0.177 AU and 0.2. The planet was treated as a test particle. The cross-hair marks the best-fit osculating elements $(a,e)=(0.64~\textnormal{AU},0.1)$ of the planet. Arrows indicate the location of mean-motion resonances. Yellow (or light) color denotes chaotic (possibly unstable) orbits and blue (dark) color denotes quasi-periodic orbits with $|\langle Y\rangle - 2.0| \simeq 0.001$. Lower panel: Same as the upper panel but now zooming into a narrower region of $(a,e)$-space of the planet. \emph{See electronic version for colors}.}
\label{fig:Map002}
\end{figure}

For reasons of optimization and execution speed we applied the most recent implementation of the MEGNO technique using Poincar\'e variables \citep{Gozdziewski2003, Gozdziewski2008}. In this reference frame the coordinates of each body are expressed in an astrocentric frame with the most massive object at rest and velocities are relative to the systems barycenter. We use geometric Jacobian coordinates as our initial conditions where the planets orbital elements are relative to the barycenter of the binary system. The Jacobian cartesian coordinates are then transformed to Poincare elements. The numerical integration of the equations of motion and the corresponding variational equations \citep{MikkolaInnanen1999} are based on the ODEX\footnote{http://www.unige.ch/math/folks/hairer/} integration algorithm \citep{hairer1993}.

To compute MEGNO maps we used the MECHANIC\footnote{http://git.astri.umk.pl/projects/mechanic.} software \citep{Slonina2012a,Slonina2012b} which utilizes a Message Passing Interface (MPI) based framework that implements massive single serial dynamical problems for parallel execution in a multi-processor computing environment. Usually we allocated 60 CPUs to compute a single dynamical MEGNO map with typical resolution of 500 x 350 initial conditions. Typical integration time of a single orbit covered about $5 \times 10^4$ binary orbits ($\simeq 2700$ years).

When investigating a dynamical system the MEGNO factor (usually denoted as $\langle Y \rangle)$ has the following mathematical properties. In general, MEGNO has the parameterization $\langle Y \rangle(t) = \alpha t + \beta$ (see references above). For a quasi-periodic chaotic initial condition, we have $\alpha \simeq 0$ and $\beta \simeq 2.0$ for $t \rightarrow \infty$ asymptotically. If the orbit is chaotic, then $\langle Y\rangle \rightarrow \lambda t/2$ for $t \rightarrow \infty$. Here $\lambda$ is the maximum Lyapunov exponent of the considered initial condition. In practice, we terminate a given numerical integration of a chaotic orbit when $\langle Y\rangle > 12$. In our MEGNO maps we color code chaotic initial conditions with yellow and quasi-periodic with blue color.

In Fig.~\ref{fig:Map002} we show a MEGNO map displaying the dynamics of a planet in semi-major axis $(a)$ and eccentricity $(e)$ space. The planet was assumed (justified by the planet displaying transit-like signals) to be co-planar with the binary orbit and all planetary orbital angles were set initially to zero. We remind the reader that the shown elements are referred to a Jacobian reference system with the planets semi-major axis being measured relative to the binary barycenter. The secondary binary component was also started with all orbital angles initially set to zero. The mass of the two binary components were set to 1.47 and $0.37~M_{\odot}$ for the primary and secondary component respectively. Referring to Fig.~\ref{fig:Map002} we identify the location of several mean-motion resonances in the system appearing as vertical chaotic columns at constant semi-major axis. The derived orbital elements of the circumbinary planet have $(a,e)=(0.64,0.1)$ placing the planet in a quasi-periodic region between the 6:1 and 7:1 mean-motion resonance. We show a zoom-in plot in the lower panel of Fig.~\ref{fig:Map002}. For planetary eccentricities larger than $e \simeq 0.25$ the planets orbital pericenter distance $(q=a(1-e))$ starts to become comparable to the binary separation resulting in close encounters and hence strongly chaotic and most possibly unstable behavior. We detected collisions of the test mass with one of the binary components and/or ejection of the planet from the system by carrying out single orbit integrations of the planetary orbit for eccentricities around 0.25 and higher.

We investigated the significance of the mass parameter of all involved objects on the overall dynamics. We recomputed the map shown in Fig.~\ref{fig:Map002} for various masses of the planet by considering $1~M_{earth}, 10~M_{earth}, 1~M_{jup}, 10~M_{jup}$ and $50~M_{jup}$ masses and found no significant qualitative difference in the global dynamics of Fig.~\ref{fig:Map002}. From comparison we found that all maps resemble each other more or less qualitatively. In a similar mass parameter survey study we also considered different mass combinations of the binary components that seemed most plausible (see previous sections) and conclude that only small qualitative changes were observed for small changes in the components mass. However, the most significant changes were detected to be close to the location of mean-motion resonances. We note that the timescales of precession frequencies of some of the orbital elements might change significantly for different mass parameters. In addition, we also note that the ETVs can also change significantly for different masses as discussed earlier. 

We conclude this section by noting that the best-fit orbital parameters locates the transiting planet KIC4862625b in a quasi-periodic region in $(a,e)$-space rendering our solution to be plausible from a dynamical point of view. We point out that our work suggest that the planet is relatively close to the 7:1 mean-motion resonance with the binary. We speculate that future observations will reveal whether the planet is or is not part of a resonant configuration.
\section{Discussion}
\label{sec:discussion}

A single transit event indicates the presence of a third
body in an eclipsing binary system. Two transits can define an approximate orbital period for
the third body, albeit with uncertainties associated with aliasing and data gaps, or orbital
inclination that can prevent a planet from transiting the moving targets of the two stars on
each passage of the planet. Even with a few planetary transits observed, unraveling the system configuration
can be challenging.

In general order of increasing challenge, the system configurations for circumbinary planets
can be solved for double-lined, double-dipped circumbinaries (planet
transits are detected across both stars, e.g. Kepler 34 and 35), followed
by double-lined, single-dipped circumbinaries (planetary transits are detected across only one
of the stars) and the single-lined circumbinaries, double-dipped and single-dipped.
For double-lined eclipsing binary stars, spectra provide the individual stars'
masses; for single-lined eclipsing binaries, spectra provide only
the mass function that relates the two masses to each other, not the two masses individually.
Transits of a third body can break the latter degeneracy inherent to single-lined systems.
Single-lined, double-dipped systems wherein transits across both stars
occur during a single binary orbit, allow excellent constraints on the
masses of the binary stars, as anticipated by Schneider \& Chevreton (1990) and demonstrated for Kepler-16
by Doyle et al. (2011).
Whether a binary is single- or double-lined depends on observational capabilities; for example, 
Kepler-16 was originally a single-lined double-dipped system; it
has since changed its category to a double-lined system due to very
high sensitivity spectra (Bender et al. 2012).
Likewise, whether a system is single- or double-dipped
also depends on the observational capabilities: even if a planet transits both stars, we may not be
able to discern the transits of the fainter star, e.g. \object{KIC 4862625}. 
More challenging than the systems that are either double-lined or double-dipped are the
single-lined and single-dipped circumbinary systems such as Kepler-38, Kepler-47 and \object{KIC 4862625}.
A large number of planetary transits, as in the case of Kepler-47b,
helps photodynamical modeling to constrain system parameters via Eq. \ref{eq:durations}
(cf. Figure \ref{fig:dur_K47b}). However, if only a few
transits occur, e.g. Kepler-38 and \object{KIC 4862625}, there may be doubt as to the
uniqueness of a solution with a large number of system parameters.

From the broadening of the spectral lines and the period of amplitude modulations
in the light curve, we infer rotation of star A and determine its
radius, which in combination with $log(g)$ from stellar spectra or the density of star A from the stellar eclipses,
either one indicates
a Solar-mass primary star. Small- or undetectable deviations from a
linear ephemeris in the primary eclipse times prove the planet is of planetary, not stellar,
mass.

The nine planetary transits further constrain the parameters of the two
stars, as their center times and durations depend on the ratio of
the stellar masses and on the transiting chord length (Eq. \ref{eq:durations}).
The latter breaks the degeneracy inherent to the mass function of the single-lined spectra.
Thereby we constrain the individual stellar masses of
of \object{KIC 4862625} and Star A's radius, assuming central planetary transits.
Similarly, we confirm the parameters of the Kepler-47 system reported by Orosz et al. (2012).

To solve the dynamical nature of the \object{KIC 4862625} system we did a
minimization search over a grid space defined by $m_1$, $a_3$,
$e_3$, and $\omega_3$, using a three-body numerical integrator in
Jacobian coordinates. Taking the first event as a reference point,
we found a set of best-fit solutions, defined by the system parameters
of the four-dimensional grid space listed above, that predict the midtransit times and durations of
the subsequent four events to within an hour. The simulations match
the observed events well but we caution that the combination
of fixed grid resolution, triaged parameters space, and the small
number of transits limit our ability to choose one of
the best fit models over another. Observations of a few additional transits
will differentiate our models, because the optimal
solutions diverge in their predictions for the center
times of the planetary transits. 

To detect chaotic solutions in the parameter
space, we studied the long term stability of the system using an
extensive set of numerical simulations, applying the MEGNO factor.
We tested systems with different planetary masses, between $1~M_{earth}$
and $50~M_{jup}$, to evaluate the changes in the dynamical behavior
of the three bodies. We do not detect significant difference outside
mean motion resonances; a planet starting near a mean-motion
resonance, however, exhibits erratic behavior. The ratio of binary
star period to the period of the giant planet is, however, not an
integer value, giving us confidence that the planet is not on
a chaotic orbit. Its orbit is near but beyond the instability region;
the ratio of planetary to binary semi-major
axis is $\sim 3.6$, compared to $\sim 2.8$ of Holman \& Wiegert (1999),
which is similar to the other Kepler planets and in agreement
with theoretical predictions that such configurations should be
common.
\section{Conclusions}
\label{sec:conclusions}

We report the discovery and characterization of a gas giant 
$r = \vRplanet \pm \eRplanet\ r_{Jupiter}$ circumbinary
planet transiting the \object{KIC 4862625} eclipsing binary system. The planet revolves around
the two stars every ${\sim}138$ days and transits the $\vMstarA \pm \eMstarA\ \Msun$ and 
$\vRstarA \pm \eRstarA\ \Rsun$ F8 primary on a $0.64$ AU, $e=0.1$ orbit.
We describe a simple, semi-automatic procedure specifically designed to search for aperiodic
transits in the light curves of binary stars. 
After identifying the transiting signals, we obtained
spectra of the two pairs of binary stars, from which we determined
the mass function, eccentricity and argument of periapsis for each pair.

We also describe the independent discovery of
the Kepler-47bc circumbinary planets by the same method. We discontinued
our characterization of that system soon after it was reported by Orosz et al. (2012).
Our truncated characterization confirms the parameters they reported.

We coin a phrase to describe circumbinary planetary systems: if planetary
transits are detected for only one star, the system is ``single-dipped,''
and for both stars, ``double-dipped.'' We discuss the relative merits of
single- or double-lined and single- or double-dipped circumbinary systems.

Periodic variations in the radial velocity of star A measured by Doppler boosting
compare favorably with those obtained by the traditional spectroscopic methods (Figure \ref{fig:boosting}).
The example of \object{KIC 4862625} demonstrates the potential of the Doppler-boosting technique for
reconnaissance of eclipsing binary stars prior to, or in lieu of, obtaining high-resolution spectra of them.

The family portrait of circumbinary planets discovered by the Kepler
mission is filling up quickly, with now seven planets reported
in less than a year since the first was reported by Doyle et al. (2011). 
\object{KIC 4862625} joins its peers Kepler
16b, 34b and 35b in the category of gas giant planets and, like Kepler
38b, orbits a binary system that includes
an evolved primary star. With the continued operation of the
Kepler mission and its exquisite-quality data,
we expect the discovery of circumbinary planets to continue
and give us a deeper insight into these exciting systems.
\acknowledgments

The authors gratefully acknowledge everyone who has contributed
to the Kepler Mission, which is funded by NASA's Science Mission Directorate. 
The \object{KIC 4862625} system has been identified and studied independently by Schwamb et al. (2012); we especially
acknowledge the collegiality of Meg Schwamb and Debra Fischer.
We acknowledge conversations with 
Josh Carter,
Nicolas Crouzet,
Selma De Mink,
Holland Ford,
Danny Lennon,
Douglas Long,
David Neufeld,
Colin Norman,
Jerome Orosz,
Rachel Osten,
M. S{\l{}}onina,
and
K. Go{\'z}dziewski.

This research used observations made with 1) the SOPHIE instrument on the 1.93-m telescope at Observatoire de Haute-Provence (CNRS), France, as part of program 12B.PNP.MOUT and 2) the Dual-Imaging Spectrograph on 3.5-m Apache Point Telescope, as part of programs Q2JH01 and Q2JH07.
This research made use of the 
the SIMBAD database, operated at CDS, Strasbourg, France;
data products from the Two Micron All Sky Survey (2MASS),
the Digitized Sky Survey (DSS),
the NASA exoplanet archive NexSci\footnote{http://exoplanetarchive.ipac.caltech.edu};
the Exoplanet Data Explorer\footnote{http://exoplanets.org} of Wright et al. (2011);
source code for transit light curves (Mandel \& Agol 2002);
an eclipsing binary simulator\footnote{http://astro.unl.edu/naap/ebs/animations/ebs.html} ;
the theoretical models\footnote{http://svo.cab.inta-csic.es/theory/db2vo/index.php} of Hauschildt et al. (1999);
a library\footnote{http://www.eso.org/sci/observing/tools/standards/IR\_spectral\_library.html}
of stellar spectra of Pickles (1998); 
SFI/HEA Irish Centre for High-End Computing (ICHEC);
PLUTO computing cluster at KASI.
V.B.K. and P.R.M. received funding from NASA Origins of Solar Systems grant NNX10AG30G and from HST DD grant 11945. 
T.C.H acknowledges support by the Korea Research Council for Science and Technology (KRCF) through the Young Scientist Research Fellowship Program grant number 2012-1-410-02 and the SFI/HEA Irish Centre for High-End Computing (ICHEC) for the provision of computational facilities and
support. This research was performed in partial fulfillment of the requirements of the PhD of V.B.K. at Johns Hopkins University.
\newpage
 
\begin{table}
\begin{center}
\footnotesize
\caption{Measured radial velocities.}
\label{tab:RadVel}
\begin{tabular}{lrrr}
\hline
\hline
BJD$_{\rm UTC}$ & RV & $\pm$$1\,\sigma$ &  Telescope/\\
$-2\,400\,000$ & (km\,s$^{-1}$) & (km\,s$^{-1}$) & Instrument \\
\hline
\multicolumn{3}{l}{\textbf{\hspace{0.7cm} KIC 4862625}}  \\
56\,029.9593                    &       $-30.3$ &       4.1  &  APO \\
56\,093.8661                    &       $-35.7$ &       4.1  &  APO \\
56\,097.8526                    &       $-16.3$ &       4.4  &  APO \\
56\,111.6009                    &       $-41.0$ &       3.6  &  APO \\
56\,159.5657                    &       $-8.36$ &       0.45    & OHP193/SOPHIE \\
56\,162.5848                    &       $-7.95$ &       0.64    & OHP193/SOPHIE \\
56\,175.3481                    &       $-28.62$        &       0.76    & OHP193/SOPHIE \\
56\,176.3531                    &       $-20.94$        &       0.62    & OHP193/SOPHIE \\
56\,177.2970$^\dagger$  &       $-15.64$        &       0.50    & OHP193/SOPHIE \\
56\,185.5063                    &       $-16.49$        &       0.49    & OHP193/SOPHIE \\
56\,188.4707                    &       $-29.65$        &       0.64    & OHP193/SOPHIE \\
56\,193.5113                    &       $-45.37$        &       1.03    & OHP193/SOPHIE \\
\hline
\multicolumn{3}{l}{\textbf{\hspace{0.7cm} KIC 10020423 (Kepler-47)}}  \\
56\,029.9593                    &       $-65.9$ &       4.0             &  APO$^{\dagger\dagger}$ \\
56\,093.8661                    &       0.0             &       4.5             &  APO \\
56\,097.8526                    &       $-59.7$ &       5.1             &  APO \\
56\,159.5866                    &       26.67   &       0.08            & OHP193/SOPHIE \\
56\,160.5787                    &       34.88   &       0.13            & OHP193/SOPHIE \\
56\,161.5490                    &       24.22   &       0.08            & OHP193/SOPHIE \\
56\,178.3126                    &       $-19.07$        &       0.16            & OHP193/SOPHIE \\
\hline
\multicolumn{4}{l}{$\dagger$: measurement corrected for sky background pollution.} \\
\multicolumn{4}{l}{$\dagger\dagger$: not absolute to barycentric reference} \\
\end{tabular}
\end{center}
\end{table}

\clearpage
\begin{table}[ht]
\begin{center}
\footnotesize
\caption{Parameters of the Binary Star Systems.
\label{tab:parameters}}
\begin{tabular}{llllll} 
\hline
\hline
{\bf KIC 4862625} & & & & &  \\
\hline
Parameter & Symbol & Value & Uncertainty (1$\sigma$) & Unit & Note \\
\hline
Orbital Period & P & 20.000214 & - & d & Pr{\v s}a et al. (2011) \\
Epoch of primary eclipse & $T_{transit}$ & 2454967.81963 & - & BJD & Pr{\v s}a et al. (2011) \\
Epoch of secondary eclipse & $T_{occultation}$ & 2454975.6738 & 0.001 & BJD &  \\
Epoch of Periastron passage & $T_{0}$ & 2454973.862 & 0.15 & BJD &  \\
Velocity semi-amplitude & $K_1$ & 19.76 & 0.73 & km s$^{-1}$ & \\
Velocity offset & $\gamma(SOPHIE)$ & -23.38 & 0.38 & km s$^{-1}$ & \\
Velocity offset & $\gamma(APO)$ & -20.7 & 1.8 & km s$^{-1}$ & \\
Argument of Periapsis & $\omega$ & 214.3 & 5.4 & \arcdeg & \\
Eccentricity   & e & 0.204 & 0.012 & & \\
Semimajor Axis & a & 0.1769 & 0.005 & AU & \\
Orbital Inclination & $i$ & 87.59 & 0.006 & \arcdeg & \\
Normalized Semimajor Axis & $a/r_A$ & 22.08 & 0.15 & & \\
Radius of Star A & $r_A$ & \vRstarA & \eRstarA & R$_\odot$ & \\
Radius of Star B & $r_B$ & \vRstarB & \eRstarB & R$_\odot$ & \\
Ratio of radii & $k$ & 0.201 & 0.015 & & $k = r_B/r_A$\\
Mass of Star A & $m_A$ & \vMstarA & \eMstarA & M$_\odot$ & \\
Mass of Star B & $m_B$ & \vMstarB & \eMstarB & M$_\odot$ & \\
Temperature of Star A & $T_A$ & 6200 & 150 & K & \\
Temperature of Star B & $T_B$ & 3390 & 50 & K & \\
V sin i of Star A & $v sin i$ & 31 & 2 & km s$^{-1}$ & \\
Fe/H of Star A & $[Fe/H]$ & -0.15 & & & \\
Gravity of Star A & log(g) & 4.0 & 0.2 & & Spectroscopic \\
Isochronal age of Star A & & 2.6 & $+3.6/-0.3$& Gyr & \\
\hline
{\bf KIC 10020423 (Kepler-47)} & & & & &  \\
\hline
Orbital Period & P & 7.44837695 & 0.00000021 & d & Orosz et al. (2012) \\
Epoch of primary eclipse & $T_{transit}$ & 2454963.24539 & 0.000041 & BJD & Orosz et al. (2012) \\
Epoch of secondary eclipse & $T_{occultation}$ & 2454959.426986 & 0.000277 & BJD & Orosz et al. (2012) \\
Epoch of Periastron passage & $T_{0}$ & 2454965.792 & 0.076 & BJD &  \\
Velocity semi-amplitude & $K_1$ & 31.18 & 0.12 & km s$^{-1}$ & \\
Velocity offset & $\gamma(SOPHIE)$ & 4.34 & 0.09 & km s$^{-1}$ & \\
Velocity offset & $\gamma(APO)$ & -32.2 & 2.8 & km s$^{-1}$ & \\
Argument of Periapsis & $\omega$ & 215.4 & 3.7 & \arcdeg & \\
Eccentricity & e & 0.0244 & 0.001 & & \\
\hline
\hline
\end{tabular}
\end{center}
\end{table}

\clearpage
\begin{table}[ht]
\begin{center}
\footnotesize
\caption{Best-fit Parameters for the Dynamical Model.
\label{tab:planmods}}
\begin{tabular}{ll}
\hline
\hline
{\bf Binary Stars} & Model 1 \\
\hline
Mass of Primary Star [\Msun] & \vMstarA \\
Mass of Secondary Star [\Msun] & \vMstarB \\
Radius of Primary Star [\Rsun] & \vRstarA \\
Radius of Secondary Star [\Rsun] & \vRstarB \\
Gravity of Star A [log(g)] & 4.14 \\
Gravity of Star B [log(g)] & 4.94 \\
Semimajor Axis [AU] & 0.1769 \\
\hline
{\bf Circumbinary Planet} & \\
\hline
Orbital Period [days] & 138.51 \\
Semimajor Axis [AU] & 0.642 \\
Eccentricity & 0.1 \\
Argument of Periastron [degrees] & 105 \\
Orbital Inclination [degrees] & 90 \\
\hline
\hline
\end{tabular}
\end{center}
\end{table}

\clearpage
\begin{table}[ht]
\begin{center}
\scriptsize
\caption{Parameters of observed and predicted planetary transits.
\label{tab:future}}
\begin{tabular}{lllllll|ll}
\hline
\hline
 Event \# & Center & $\sigma$ & Depth$^\dagger$ & $\sigma$ & Duration & $\sigma$ & Center & Duration \\
 & (BJD-2450000) & (Center) & [ppm] & (Depth) & [days] & (Duration) & [BJD-2450000] & (days) \\
\hline
\hline
{\bf Observed} & & & & & & & {\bf Model} 1 & \\
\hline
1 & 5070.8267 & 0.019 & 870 & 90 & 0.5485 & 0.0378 & --- & 0.56 \\
2 & 5207.4077 & 0.011 & 631 & 90 & 0.5125 & 0.021 & 5207.43 & 0.51 \\
3 & 5344.1308 & 0.012 & 914 & 83 & 0.6184 & 0.023 & 5344.14 & 0.55 \\
4 & 5480.004 & 0.015 & 1042 & 78 & 0.7513 & 0.029 & 5479.99 & 0.76 \\
5 & 5613.2329 & 0.013 & 1043 & 67 & 0.891 & 0.032 & 5613.22 & 0.84 \\
6 & 5749.1914 & 0.013 & 939 & 95 & 0.4680 & 0.043 & 5749.25 & 0.5 \\
7 & 5885.9215 & 0.035 & 1192 & 98 & 0.5235 & 0.074 & 5885.97 & 0.54 \\
8 & 6022.3334 & 0.037 & 974 & 46 & 0.5787 & 0.036 & 6022.36 & 0.57 \\
9 & 6157.0322 & 0.02 & 1559 & 87 & 1.1353 & 0.034 & 6157.00 & 1.14 \\
\hline
{\bf Future} & & & & & & \\
\hline
 10 & --- & --- & --- & --- & --- & --- & 6291.04 & 0.61 \\ 
 11 & --- & --- & --- & --- & --- & --- & 6427.60 & 0.54 \\ 
\hline
\hline
\multicolumn{4}{l}{$\dagger$: in terms of $(\frac{r_p}{r_A})^2$} \\
\end{tabular}
\end{center}
\end{table}
\end{document}